\documentclass[12pt]{iopart}
\usepackage{iopams}
\usepackage{color,graphicx}
\usepackage[colorlinks,pdfstartview=FitH,pdfpagemode=UseNone,pdfpagelayout=SinglePage,bookmarksopen=false]{hyperref}

\begin{document}

\title[Dynamics of RSAE and EPs during current ramp-up]{Dynamics of reversed shear Alfv\'en eigenmode and energetic particles during current ramp-up}

\author{Tao Wang$^{1,2,3}$, Zhiyong Qiu$^3$, Fulvio Zonca$^{4,3}$, Sergio Briguglio$^4$ and Gregorio Vlad$^4$}

\address{$^1$\quad Key Laboratory of Optoelectronic Devices and Systems of Ministry of Education and Guangdong Province, College of Physics and Optoelectronic Engineering, Shenzhen University, Shenzhen 518060, People's Republic of China\\
$^2$\quad Advanced Energy Research Center, Shenzhen University, Shenzhen 518060, People's Republic of China\\
$^3$\quad Institute for Fusion Theory and Simulation and Department of Physics, Zhejiang University, Hangzhou 310027, People’s Republic of China\\
$^4$\quad ENEA, Fusion and Nuclear Safety Department, C. R. Frascati, Via E. Fermi 45, 00044 Frascati (Roma), Italy}

\ead{t\_wang@zju.edu.cn}

\begin{abstract}
Hybrid MHD-gyrokinetic code simulations are used to investigate the dynamics of frequency sweeping reversed shear Alfv\'en eigenmode (RSAE) strongly driven by energetic particles (EPs) during plasma current ramp-up in a conventional tokamak configuration.
A series of weakly reversed shear equilibria representing time slices of long timescale MHD equilibrium evolution is considered, where the self-consistent RSAE-EP resonant interactions on the short timescale are analyzed in detail.
Both linear and nonlinear RSAE dynamics are shown to be subject to the non-perturbative effect of EPs by maximizing wave-EP power transfer.
In linear stage, EPs induce evident mode structure and frequency shifts; meanwhile, RSAE saturates by radial decoupling with resonant EPs due to weak magnetic shear, and gives rise to global EP convective transport and non-adiabatic frequency chirping.
The spatiotemporal scales of phase space wave-EP interactions are characterized by the perpendicular wavelength and wave-particle trapping time.
The simulations provide insights into general as well as specific features of RSAE spectra and EP transport from experimental observations, and illustrate the fundamental physics of wave-EP resonant interaction with the interplay of magnetic geometry, plasma non-uniformity and non-perturbative EPs.
\end{abstract}


\submitto{\NF}

\maketitle

\section{Introduction}
\label{sec:introduction}

In tokamak plasmas of fusion interest, the energetic/fast particles (EPs), generated by nuclear fusion reactions and/or external power inputs such as neutral beam injection (NBI) or ion cyclotron resonance heating (ICRH), play crucial roles in plasma heating and current drive; and thus, must be well confined to achieve better performance.
An important mechanism of anomalous EP transport is via collective resonant excitation of shear Alfv\'en wave (SAW) fluctuations \cite{chen_rmp16}, in the form of Alfv\'en eigenmodes (AEs) \cite{cheng_ap85} or energetic particle continuum modes (EPMs) \cite{chen_pop94}.
If strongly driven, these fluctuations could induce substantial power loss and localized thermal flux onto plasma facing components by EP transport before thermalization \cite{heidbrink_nf91,duong_nf93}.
During the past several decades, significant achievements in comprehending the basic and general physics aspects on SAW-EP dynamics have been greatly gained by intense experimental and theoretical researches \cite{heidbrink_nf94,wong_ppcf99,vlad_rnc99,pinches_ppcf04b,heidbrink_pop08,breizman_ppcf11,lauber_pr13,gorelenkov_nf14,chen_rmp16,todo_rmpp19}.
Nevertheless, more dedicated efforts are needed to incorporate multiple physics ingredients and address situations of practical interest; thereby, validating and deepening the present understanding, and envisioning burning plasma experiments in the near future \cite{heidbrink_pop02,zonca_ppcf06b,fasoli_nf07,chen_nf07b,pinches_pop15,chen_rmp16}.

In this work, connected with both fundamental physics and realistic experiments, we investigate, via numerical simulations, linear and nonlinear SAW-EP dynamics during the plasma current ramp-up phase in a conventional tokamak discharge.
In particular, we consider a generalized scenario where high external power is applied to a plasma with relatively low density and current.
Due to insufficient current penetration and/or diffusion, a weakly reversed shear magnetic configuration is usually created, where the radial profile of safety factor $q$ contains a local minimum, $q_{\rm min}$, which typically resides deeply in the plasma core region.
Concurrently, an anisotropic and concentrated population of fast ions is also produced, and could readily excite the reversed shear AE (RSAE, also dubbed Alfv\'en cascade) \cite{sharapov_pla01,berk_prl01}, in addition to the more common toroial AE (TAE) \cite{cheng_ap85,cheng_pf86,wong_prl91,heidbrink_nf91}.
In fact, RSAE is empirically recognized as a prominent signature of reversed shear plasmas by its characteristic sweeping frequency, which follows temporal evolution of $q_{\rm min}$ with a quasi-coherent spectral line for each toroidal mode number $n$ \cite{sharapov_pop02} (see \cite{breizman_ppcf11} for a recent review).
In most representative cases, when $q_{\rm min}$ decreases from a rational value $m/n$ to $(m-1/2)/n$, RSAE exhibits upward sweeping frequency up to TAE frequency band, $\omega_{\rm TAE}\simeq v_{\rm A}/(2qR)$.
Here, $m$ is the (dominant) poloidal mode number, $v_{\rm A}$ is Alfv\'en speed, and $R$ is major radius.
Indeed, upward frequency sweeping RSAEs with ramping-up plasma current and continuous NBI and/or ICRH have been routinely observed in many devices \cite{kimura_nf98,sharapov_pla01,snipes_pop05,vanzeeland_prl06,garcia-munoz_nf11}.
Of particular importance, the fluctuations dominated by RSAE and TAE often cause significant anomalous fast ion transport/losses, as both short timescale intermittent convective pulses \cite{takechi_pop05,garcia-munoz_prl10,pace_ppcf11} and accumulative profile flattening on long timescale \cite{ishikawa_nf06,heidbrink_nf08,vanzeeland_pop11}.
Motivated by the rich phenomenology of SAW-EP interactions and the importance of unveiling the underlying physics mechanism, many dedicated numerical simulation works \cite{vlad_nf09,white_ppcf10,tobias_prl11,deng_nf12b,schneller_nf13,todo_nf14,zhangruibin_nf15,podesta_nf16} have been devoted to analyzing various aspects of the observations.
However, previous works either focus on short timescale dynamics with fixed magnetohydrodynamic (MHD) equilibrium profiles; or sacrifice crucial self-consistent description of SAW-EP interaction to extend the simulation time span.
Accordingly, on one hand, in order to fully capture the frequency sweeping feature of RSAE, it is necessary to take into account the finite equilibrium evolution in simulation setup.
On the other hand, a self-consistent, non-perturbative description of SAW-EP interactions is also crucially needed, since the non-perturbative effect of EPs plays a significant role in RSAE dynamics when the driven fluctuation deviates from instability threshold \cite{berk_prl01,zonca_pop02,wangtao_pop18,wangtao_pop19}.
Here, the non-perturbative effect corresponds to the important contribution of EPs in determining mode structure and real frequency on both linear and nonlinear dynamics, such that SAW-EP resonance condition is best maintained and corresponding power transfer is maximized \cite{chen_prl84,tsai_pfb93,chen_pop94,zonca_pop96,briguglio_pop98,zonca_nf05,zonca_pop14a,zonca_pop14b,zonca_njp15,chen_rmp16,wangtao_pop18,wangtao_pop19}.
Indeed, the deviation of RSAE spectra from MHD limit has been identified in experiments with high power input \cite{kusama_nf99,shinohara_nf01,shinohara_ppcf04,pinches_ppcf04b,sharapov_nf06,ishikawa_nf06,tobias_prl11}, suggesting the important role of non-perturbative EPs.
More specifically, rapid and repetitive frequency chirping within $1~{\rm ms}$, i.e., characteristic timescale of a few inverse linear growth rates \cite{zonca_iaea02,zonca_nf05}, could take place in addition to much slower frequency sweeping in the timescale of $\Or(100~{\rm ms})$ induced by equilibrium evolution.
The phenomenon of multi-timescale RSAE frequency chirping and sweeping is addressed by the simulations presented in this paper, where we specify ``chirping'' as fast, short timescale and ``sweeping'' as slow, long timescale frequency evolution, in the sense described above.

\begin{figure}
\includegraphics{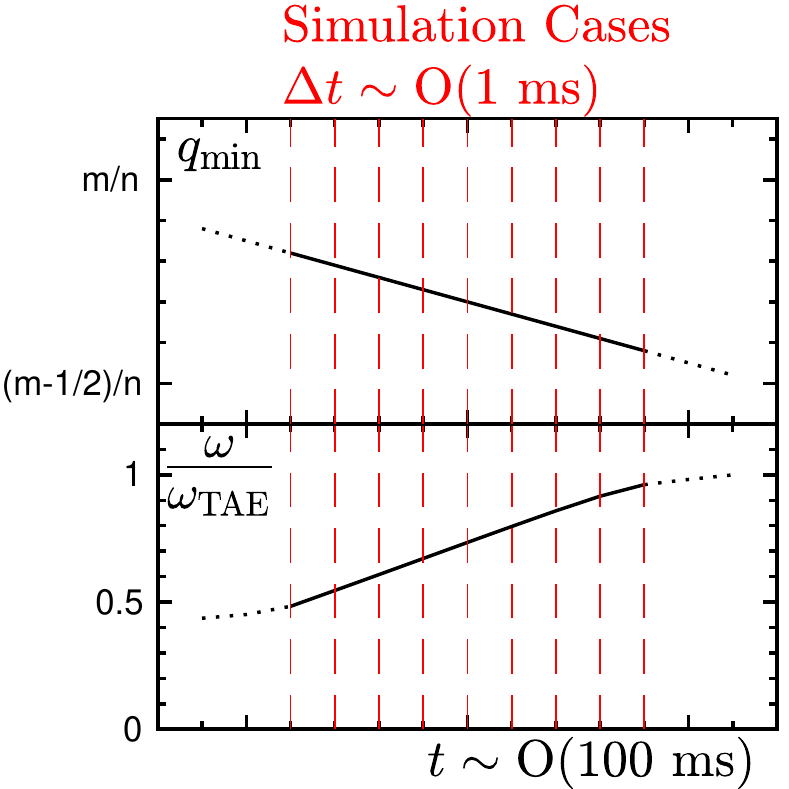}
\caption{A schematic viewgraph of the simulation approach and relevant timescales.}
\label{fig:viewgraph}
\end{figure}

In order to be feasible within the limitation of computational resources and numerical model assumptions, we utilize the timescale separation between fast SAW-EP interaction and slow equilibrium evolution, and set up a series of \emph{ad hoc} simulations to investigate RSAE-EP dynamics with superimposed small but finite equilibrium change.
To be more precise, several self-similar MHD equilibria representing time slices of ramping-up plasma current in long timescale are used as the only changing input variable for multiple simulation cases, in which the short timescale nonlinear SAW-EP interaction is described self-consistently and in which the equilibrium is kept fixed within each simulation case.
A schematic viewgraph of the simulation approach and relevant timescales is shown in figure~\ref{fig:viewgraph}.
Admittedly, this approach is not fully self-consistent.
In particular, we note that in order to isolate the effect of small equilibrium changes, the initialized EP  distribution function is effectively unchanged among cases.
In other words, the slow nonlinear equilibrium variation of EP distribution under the effects of fast ion source/sink, Coulomb collision, micro-turbulence and SAW-EP nonlinear interaction \cite{zonca_nf05,chen_nf07a,zonca_njp15,falessi_pop19a} is not taken into account in the scope of this work.
These effects can be recovered within the theoretical framework of EP phase space zonal structure transport \cite{zonca_njp15,chen_rmp16,falessi_pop19a,falessi_njp20}, and will be the subject of a future work.
Nevertheless, the simulations illustrate fast frequency chirping and EP convective transport on short timescales; and present the effect of long timescale equilibrium variation via transparent inter-case comparison.

With the intention of analyzing the general and specific features of RSAE-EP dynamics observed in various devices in mind, the simulation setup is necessarily simplified while representative, and includes all crucial physics ingredients such as magnetic geometry, plasma non-uniformity and non-perturbative EPs \cite{zonca_ppcf15,zonca_njp15,chen_rmp16}.
A single toroidal mode number $n=3$ is considered; the investigated time window is selected as a signature RSAE upward frequency sweeping period up to TAE range by directly controlling $q_{\rm min}$ (see also figure~\ref{fig:viewgraph}).
Furthermore, the simulations are carried out with the hybrid MHD-gyrokinetic code (HMGC) \cite{briguglio_pop95,briguglio_pop98}, which describes self-consistent nonlinear SAW-EP interactions on short timescales with a simple but yet relevant model based on the two component fusion plasma description \cite{chen_prl84} and the hybrid pressure coupling model \cite{park_pfb92}.
Numerical details are described in section~\ref{sec:model}, along with primary simulation parameters.
In order to be relevant and realistic in the choice of parameters, the mid-size HL-2A tokamak \cite{ding_pst18} is chosen as a convenient reference whenever applicable, where the EP population is modeled as positive-ion-based neutral beam (P-NB) with a single injection pitch angle.
Note that we do not consider \emph{a priori} any specific HL-2A discharge in the present work; meanwhile, the present research can be readily extrapolated to other discharges/devices based on the dimensionless parameters which dictate the physics \cite{wangtao_pop18,wangtao_pop19}.

For the self-consistent RSAE-EP resonant interaction on short timescales, we further subdivide it into three stages by the dominant physics mechanism; namely, linear growing, nonlinear saturation, and post saturation stages.
Note that ``stage'' implies short timescale in this paper.
Main results for the three stages are presented and discussed in, respectively, sections~\ref{sec:linear}, \ref{sec:saturation}, \ref{sec:post-saturation}; where we summarize the common features of all simulation cases with varying $q_{\rm min}$, as a reflection of governing physics for the investigated scenario, and discuss inter-case differences suggesting the effect associated with slow equilibrium evolution.
On linear properties presented in section~\ref{sec:linear}, we characterize the non-perturbative effect of EPs in shaping mode structure and shifting mode frequency; meanwhile, global SAW-EP resonance structure responsible for the excited fluctuation spectra is analyzed in detail.
Both properties suggest the dominant physics mechanism in the following nonlinear dynamics; namely, fluctuation saturation, frequency chirping, and EP transport.
In the saturation stage (section~\ref{sec:saturation}), we focus on the relevant nonlinear spatiotemporal scales, which are important for understanding experimental observations as well as physics implication of simulation results.
Two reference scales are the perpendicular (to equilibrium magnetic field) wavelength, $\lambda_\perp$, related with the locality of resonant EP response; and the wave-particle trapping time, $\tau_{\rm B}$, reflecting the timescale beyond which the system response gradually becomes adiabatic.
We note that an adiabatic paradigm of nonlinear SAW-EP dynamics has been established \cite{berk_pfb90a,berk_pfb90b,berk_pfb90c,berk_prl92,breizman_ppcf11} via exploring similarities with respect to beam-plasma system \cite{bernstein_pr57,shapiro_63,mazitov_65,oneil_pf65,altshul_66,oneil_pf71}.
This ``bump-on-tail'' paradigm is generally applicable to scenarios sufficiently close to marginal stability with weak drive and strong dissipation \cite{berk_prl96,breizman_pop97,berk_ppr97,breizman_ppcf11}, and characteristic nonlinear timescale satisfying $\tau_{\rm NL}\gg\tau_{\rm B}$, i.e., adiabatic; such that EP phase space density is conserved within the separatrix between trapped and untrapped orbits, and nonlinear dynamics can be reflected by the spontaneously generated hole-clump pairs in EP phase space \cite{berk_pf70,dupree_pf72,berk_pla97,berk_pop99}.
Moreover, EP transport is implicitly local unless significant resonance overlap takes place \cite{berk_prl92,berk_nf95,berk_pop96}.
However, the adiabatic assumption breaks down for the present ``ramping'' scenario, which is dominated by high power input and consequently, by the EPs whose distribution function generally deviates from marginal stability threshold.
We demonstrate that the investigated non-adiabatic regime is characterized by rapid frequency chirping in the timescale $\tau_{\rm NL}\sim\Or(\tau_{\rm B})$ and global EP convective transport over radial scale length $\sim\Or(\lambda_\perp)$ \cite{white_pf83,briguglio_pop98,zonca_nf05,zonca_njp15,chen_rmp16}, where wave-EP phase locking could take place and give rise to convective amplification of fluctuation amplitude by maximizing power transfer \cite{zonca_nf05,zhanghuasen_prl12,vlad_nf13,bierwage_nf13,bierwage_nf14,zonca_ppcf15,zonca_njp15,chen_rmp16,vlad_njp16,bierwage_nf17a,wangtao_pop19}.
The onset of non-adiabatic frequency chirping and intrinsically nonlocal EP transport is induced by radial decoupling of resonant EPs with the non-perturbatively excited fluctuation.
The theoretical framework is outlined in \cite{zonca_njp15,chen_rmp16}, where the one-on-one correspondence of non-adiabaticity and non-perturbativity is addressed.

On application side, the present work illustrates many specific features of weakly reversed shear plasmas in typical present-day tokamaks with high power NBI heating; and thereby, extending our previous work addressing next generation devices \cite{wangtao_pop18,wangtao_pop19}.
As also discussed in \cite{wangtao_pop18,wangtao_pop19}, the plasma volume can be schematically subdivided into inner- and outer-core regions based on magnetic shear.
For radially localized RSAE in the inner-core region, the resonance structure with circulating EPs is crucially dominated by weak shear, which favors radial decoupling with resonant EP convection, and results in global EP transport associated with long wavelength and large normalized EP orbit width.
Under the effects of SAW-EP radial decoupling and non-perturbative interaction, the fluctuation splits into a ``convective'' branch maintaining phase locking with resonant EPs, and ``relaxation'' branches back to weakly damped AE states.
We note that these two branches exist in general wave-EP resonant interactions \cite{oneil_pf68}, as evidenced by experiments with strongly driven bursty fluctuations \cite{mcguire_prl83,heidbrink_ppcf95,kusama_nf99,shinohara_nf01,shinohara_ppcf04,pinches_ppcf04b,sharapov_nf05,sharapov_nf06,ishikawa_nf06,yuliming_prl20}.
Applied to non-perturbatively excited RSAEs, nonlinear dynamics is importantly regulated by magnetic geometry and plasma non-uniformity, where the downward chirping convective branch experiences stronger continuum damping \cite{zonca_prl92,rosenbluth_prl92,zonca_pfb93,zonca_pop02}, and only the relaxation branch survives in post saturation stage (section~\ref{sec:post-saturation}).
Thus, dominated by non-adiabatic frequency relaxation to MHD limit, the simulations illustrate and illuminate the observed fast frequency chirping with strong EP drive, and neatly recover the conventional MHD analysis of slow frequency sweeping.
Meanwhile, more insights into EP phase space transport are presented.
Correlated with relatively low characteristic energy of P-NBs and their significant phase space anisotropy in early phase of a discharge, on one hand, the effective resonance range in EP phase space distribution is generally broad.
On the other hand, a substantial fraction of EP drive is provided by velocity space anisotropy, in addition to the density gradient as universal instability mechanism \cite{book-chen88,fu_pfb89a}, since EP diamagnetic frequency is not much larger than mode frequency \cite{chen_rmp16}, $\omega_{\star\rm H}/\omega\sim\Or(10)$.
Consequently, significant and global EP phase space profile relaxation is induced by the fluctuations, and results in eventually decaying fluctuation amplitude after saturation.
However, EP transport is not entirely defined by linear resonance condition, due to the importance of non-adiabatic fluctuation frequency chirping, which could significantly extend phase space resonant range and intensify EP transport.
The SAW convective amplification process confirms the dictating role of EPs as maximizing SAW-EP power transfer.
Further conclusion and discussion are given in section~\ref{sec:conclusion}.

\section{Numerical model and parameters}
\label{sec:model}

\subsection{Simulation model}
\label{subsec:model}

In this work, HMGC is used to investigate the self-consistent nonlinear SAW-EP interaction via the non-perturbative pressure coupling formulation \cite{park_pfb92}.
That is, the thermal (bulk) plasmas are described by reduced MHD equations \cite{izzo_pf83} in simplified toroidal geometry characterized by shifted circular magnetic flux surfaces; EP kinetic compression enters the momentum equation via the pressure tensor term, which is computed by the corresponding gyrocenter distribution function using particle-in-cell method.
EP orbits are solved in the perturbed electromagnetic fields by nonlinear Vlasov equations in the drift-kinetic limit, i.e., finite Larmor radius effect is neglected.
The detailed model equations are presented in \cite{briguglio_pop95,vlad_pop95,vlad_rnc99,wangxin_pop11}, and are thus omitted here for simplicity.

As described above, the hybrid MHD-kinetic model \cite{park_pfb92} is a simplified but yet relevant numerical tool for the scope of present analysis, since it preserves the crucial physics ingredients such as equilibrium geometry, plasma non-uniformity and non-perturbative EP response, which are necessary to properly describe nonlinear SAW-EP interactions as articulated in \cite{zonca_ppcf15,zonca_njp15,chen_rmp16}.
Here, we note that RSAE spectrum is known to be subject to kinetic effects of thermal plasmas in the low frequency domain, including geodesic acoustic coupling and pressure gradient \cite{breizman_pop05,fu_pop06a} as well as Landau damping.
However, the kinetic extension of HMGC model \cite{wangxin_pop11,vlad_njp16} is not accounted for in this work, as we consider RSAEs near TAE frequency band in low-$\beta$ limit (cf. figure~\ref{fig:viewgraph} and section~\ref{subsec:MHD} below), where $\beta$ is the ratio of plasma thermal to magnetic pressures.
Thereby, these neglected effects will not qualitatively impact the RSAE dynamics, which is dominated by EPs.

\subsection{MHD parameters}
\label{subsec:MHD}

\begin{figure}
\includegraphics{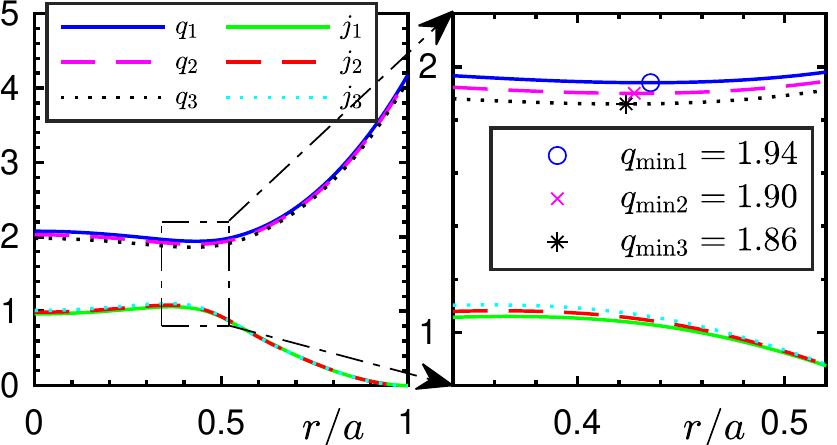}
\caption{For three reference equilibria considered in this work, the safety factor $q$ (upper curves) and current density $j$ (lower curves) profiles as functions of plasma minor radial coordinate $r$ normalized to minor radius $a$.
Here, $j$ is normalized to $B_0/R_0$.
The three equilibria labeled $1-3$ are characterized by $q_{\rm min}=1.94$, $1.90$, $1.86$, and plotted as solid, dashed and dotted curves, respectively.
The right panel is enlarged around $q_{\rm min}$, which is shown by markers.}
\label{fig:equilibrium}
\end{figure}

\begin{figure}
\includegraphics{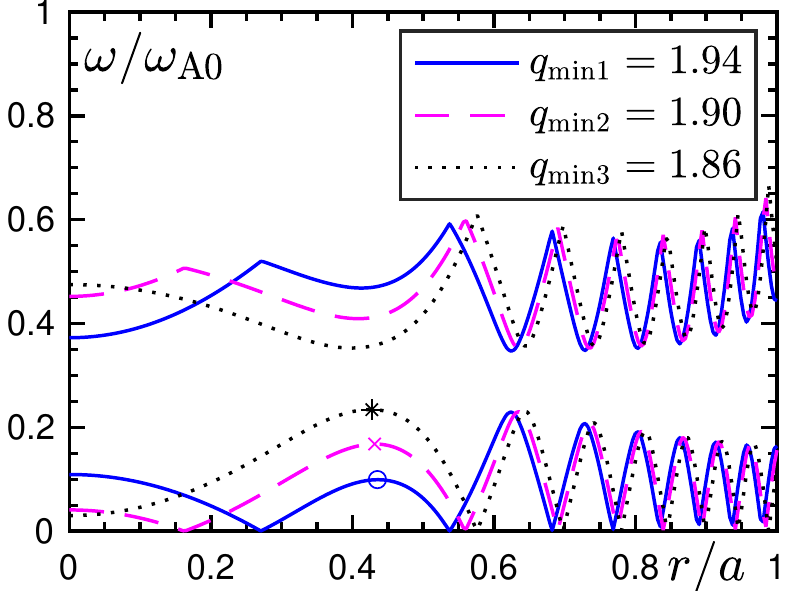}
\caption{Ideal, incompressible SAW continua of toroidal mode number $n=3$ for the three reference equilibria shown in figure~\ref{fig:equilibrium}.
The makers similar to figure~\ref{fig:equilibrium} indicate the corresponding RSAE lower continuum accumulation point.}
\label{fig:continuum}
\end{figure}

As introduced in section~\ref{sec:introduction}, simulation parameters are chosen to resemble current ramp-up phase of a present-day tokamak discharge using typical HL-2A parameters.
The plasma domain is characterized by circular poloidal cross sections with major radius $R_0=1.65~\rm m$ and minor radius $a=0.40~\rm m$.
The on-axis magnetic field is $B_0=1.3~\rm T$, where in this paper, the subscript ``0'' denotes spatially nonuniform quantities evaluated at the magnetic axis.
A series of self-similar weakly reversed $q$ profiles with $q_0\sim2$ and $q_a\sim4$ is considered in this work, where $9$ simulation cases are presented with $q_{\rm min}$ decreasing from $1.94$ to $1.86$ with a step size $\Delta q_{\rm min}=0.01$, corresponding to $\vert k_\parallel qR\vert=\vert nq-m\vert=0.18-0.42$ for the dominant $n=3$, $m=6$ perturbation at $q_{\rm min}$, with $k_\parallel$ the parallel component of wave number.
Figure~\ref{fig:equilibrium}
shows three reference MHD equilibria.
We note again that each $q$ profile is used for different cases, whilst MHD equilibrium is kept fixed within a single simulation.
As shown in figure~\ref{fig:equilibrium}, total plasma current gradually increases, where a ``bump'' in the current density profile penetrates towards the plasma core.
Such current profile evolution results in slightly decreasing $q$ profile and inward shift of $r_{q_{\rm min}}$, as one can more clearly see in the inset of figure~\ref{fig:equilibrium}.
Moreover, consistent with the reduced MHD formalism in HMGC physics model described in section~\ref{subsec:model}, thermal plasma compressibility is not considered.
Bulk ion is assumed to be Deuterium with on-axis density $n_{\rm i0}=2.0\times10^{19}~\rm m^{-3}$, and a parabolic radial profile
\begin{equation}
n_{\rm i}(r)/n_{\rm i0}=1-0.7(r/a)^2\,.
\end{equation}
Figure~\ref{fig:continuum}
shows the SAW continua for the three reference $q$ profiles.
Here, $\omega_{\rm A0}\equiv v_{\rm A0}/R_0$ is the on-axis Alfv\'en frequency used for normalization of time and frequency.
One sees that the small variation of $q$ profiles mainly alters the continuum structure in the inner-core region, where RSAE frequency increases up to TAE range as a consequence of decreasing $q_{\rm min}$.

The coupled reduced MHD equations consider two electromagnetic field variables, namely, the electrostatic potential $\phi$ and poloidal flux function $\psi$, which is related to the parallel component of magnetic vector potential $A_\parallel$ \cite{izzo_pf83,briguglio_pop95}.
HMGC uses Fourier decomposition in poloidal and toroidal directions, yielding the respective mode numbers $m$ and $n$.
As noted above, a single $n=3$ perturbation is considered, while neglecting the coupling among modes with different $n$.
The perturbed harmonics are solved by finite difference method in radial direction with a mesh $N_{\rm MHD}=256$, which is sufficient for the long wavelength fluctuations in the present work.
In addition, small values of resistivity $\eta$ and viscosity $\nu$ are included in the MHD equations for numerical stability reasons.
For all simulations considered in this paper, the resistivity corresponds to Lundquist number $S\equiv4\pi a^2\omega_{\rm A0}/(\eta c^2)=2.5\times10^5$; the normalized viscosity is $\nu\omega_{\rm A0}/a^2=10^{-8}$.
Nevertheless, ideal MHD constraint $\delta E_\parallel\simeq0$ generally applies, and we will refer to a single scalar field $\phi$ to represent the SAW fluctuations.

The mode frequency $\hat \phi(r,\omega)\vert_{t=t_{\rm ref}}$ at a reference time $t_{\rm ref}$ is obtained by fast Fourier transform (FFT) to temporal series of $\phi(r,t)$ within a finite time window; i.e., $t\in[t_{\rm ref}-\Delta t_{\rm window},~t_{\rm ref}+\Delta t_{\rm window}]$.
To minimize the influence of data prior to and after $\phi(r,t_{\rm ref})$, we weight $\phi(r,t)\vert_{t_{\rm ref}-\Delta t_{\rm window}}^{t_{\rm ref}+\Delta t_{\rm window}}$ by a Hanning function; in addition, the temporal series are padded with zeros on both sides to smooth the $\omega$ grid.
Since we are interested in the fast frequency chirping during nonlinear stage, it is necessary to choose an intermediate value of $\Delta t_{\rm window}$ in the spectral analysis \cite{bierwage_nf17a}; such that one has good frequency resolution as $\Delta\omega=\pi/\Delta t_{\rm window}$, and minimized numerical artifacts associated with the asymmetry of fluctuation amplitude (weight) within the temporal series.
In this paper, $\Delta t_{\rm window}=300~\omega_{\rm A0}^{-1}$ is used as a standard value.
In addition, we also cross-check the spectral analysis using $\Delta t_{\rm window}=100~\omega_{\rm A0}^{-1}$, to seek better confidence in the relevant timescale of frequency chirping and numerical convergence.
More details with examples are discussed in section~\ref{sec:saturation}.

\subsection{EP parameters}
\label{subsec:EP}

\begin{figure}
\includegraphics{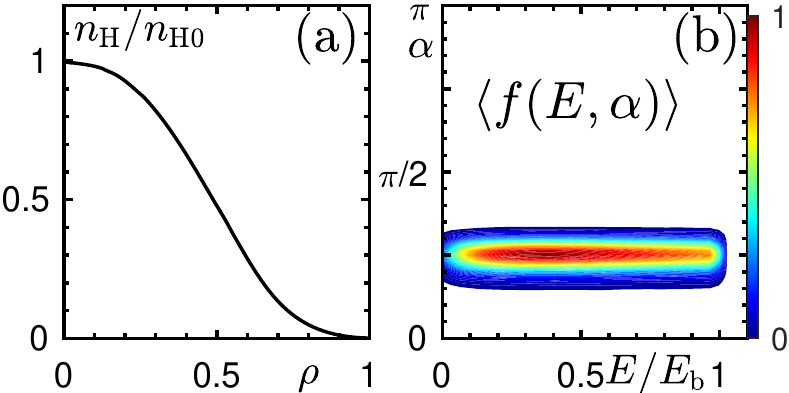}
\caption{For a reference case with $q_{\rm min}=1.90$, the EP density profile (a) and volume integrated velocity space distribution function (b) after initial relaxation.
The reported quantities are averaged over a time window in the linear stage, where the initial relaxation has completed and the fluctuation amplitude is still negligible to cause discernible EP redistribution.}
\label{fig:df90_relax}
\end{figure}

The energetic (hot) particles in the simulations are modeled as tangentially injected Deuterium P-NB ions in the co-current direction.
They are described by a model slowing-down distribution function, with birth energy $E_{\rm b}=45~\rm keV$ and uniform critical energy $E_{\rm c}=28~\rm keV$ \cite{stix_pp72}, yielding the following normalized parameters: $v_{\rm H}/v_{\rm A0}=0.327$, $\rho_{\rm H}/a=0.059$.
Here, $v_{\rm H}\equiv\sqrt{E_{\rm b}/m_{\rm H}}$ is a characteristic EP thermal velocity; $\rho_{\rm H}\equiv v_{\rm H}/\Omega_{\rm cH0}$ is the corresponding Larmor radius; and $\Omega_{\rm cH0}=e_{\rm H}B_0/(m_{\rm H}c)$ is the on-axis cyclotron frequency, with $m_{\rm H}$ and $e_{\rm H}$ the EP (Deuterium) mass and charge, respectively.
The initial distribution function reads
\begin{equation}
f_{\rm H}(\psi,E,\alpha;t=0)=\frac{3}{4\pi}\left(\frac{m_{\rm H}}{2E_{\rm c}}\right)^{3/2}
\frac{n_{\rm H}(\psi_{\rm eq})\Xi(\alpha)}{\left[(E/E_{\rm c})^{3/2}+1\right]\ln\left[(E_{\rm b}/E_{\rm c})^{3/2}+1\right]}\,.
\label{eq:df}
\end{equation}
Here, $n_{\rm H}(\psi_{\rm eq})$ is the radial density profile as a function of equilibrium flux coordinate
\begin{equation}
n_{\rm H}(\psi_{\rm eq})/n_{\rm H0}=\exp\left[-\left(\frac{\rho}{L_n}\right)^{\lambda_n}\right]\,,
\label{eq:n-psi}
\end{equation}
with $n_{\rm H0}=10^{18}~\rm m^{-3}$, $L_n=0.6$ and $\lambda_n=3$ [cf. figure~\ref{fig:df90_relax}(a)], where $\rho\equiv\sqrt{(\psi_{\rm eq}-\psi_0)/(\psi_a-\psi_0)}$ is a radial-like flux coordinate.
$\Xi(\alpha)$ is the pitch angle distribution function
\begin{equation}
\Xi(\alpha)=\frac{4}{\Delta\sqrt{\pi}}
\frac{\exp\left[-(\cos\alpha-\cos\alpha_{\rm inj})^2/\Delta^2\right]}{{\rm erf}\left[(1-\cos\alpha_{\rm inj})/\Delta\right]{\rm erf}\left[(1+\cos\alpha_{\rm inj})/\Delta\right]}\,,
\label{eq:alfa}
\end{equation}
with $\alpha\equiv\cos^{-1}(v_\parallel/\sqrt{2E/m_{\rm H}})$ the pitch angle and $v_\parallel$ the parallel velocity.
We use $\alpha_{\rm inj} = \pi/4$ and $\Delta=0.1$, i.e., the EPs are predominantly co-circulating with a relatively narrow pitch angle distribution [cf. figure~\ref{fig:df90_relax}(b)], where one can readily estimate $v_\parallel=v_{\rm H}\simeq v_{\rm A}/3$ at birth energy and injection pitch angle.

Note that the initial distribution (\ref{eq:df}) is not characterized by particles' constants of motion; in particular, the initial radial density function (\ref{eq:n-psi}) is described by flux coordinate $\psi_{\rm eq}$.
Due to the relatively large magnetic drift as $\rho_{\rm H}/a=0.059$ and $q\gtrsim2$, the initialized particles will deviate from the nominal flux surfaces and promptly relax to a genuine equilibrium distribution; thus, leading to some corrugations in the {\it de facto} loaded distribution function.
However, as the initial conditions for all simulation cases are quite similar, the largest inter-case difference in the relaxed distribution functions is $\sim\Or(10^{-3})$ (normalized to the peak value).
Since the fluctuation-induced EP transport is generally at least an order of magnitude larger than such uncertainty, we consider that all cases are initialized to the same promptly relaxed distribution \cite{vlad_nf09} as reported in figure~\ref{fig:df90_relax}.
It will be referred to as the linear distribution in later comparisons with the perturbed one due to fluctuation-induced EP transport.

The EPs coordinates are pushed in 5D phase space $(r,\theta,\zeta,\mu,v_\parallel)$ with a time step ${\rm d}t=0.06~\omega_{\rm A0}^{-1}$, where $\theta$ is the poloidal angle, $\zeta$ is the toroidal angle, and $\mu\equiv m_{\rm H}v_\perp^2/(2B)$ is the conserved magnetic moment.
The kinetic module uses $128\times128\times48$ grids in $r$, $\theta$, $\zeta$ directions, and $16$ particle markers per cell; a convergence test shows that such numerical resolution is sufficient for the present analysis.
Furthermore, EP source and Coulomb collisions are not considered in the present numerical model; meanwhile, particles that leave the simulation domain, i.e., $r/a>1$, are considered lost.
As discussed above, we focus on comparing different initial-value simulations with equilibrium profiles being the only variable in the present work.
Meanwhile, the effect of source and collisions in replenishing the phase space resonant range is not crucially important in the short timescale considered in each simulation case.
A simplified source and collision module is currently being implemented in the code, and will be reported in a future publication.
The relevance of source and collisions is further discussed in section~\ref{sec:conclusion}.

\section{Linear properties}
\label{sec:linear}

\subsection{Mode spectra}
\label{subsec:spectra}

\begin{figure}
\includegraphics{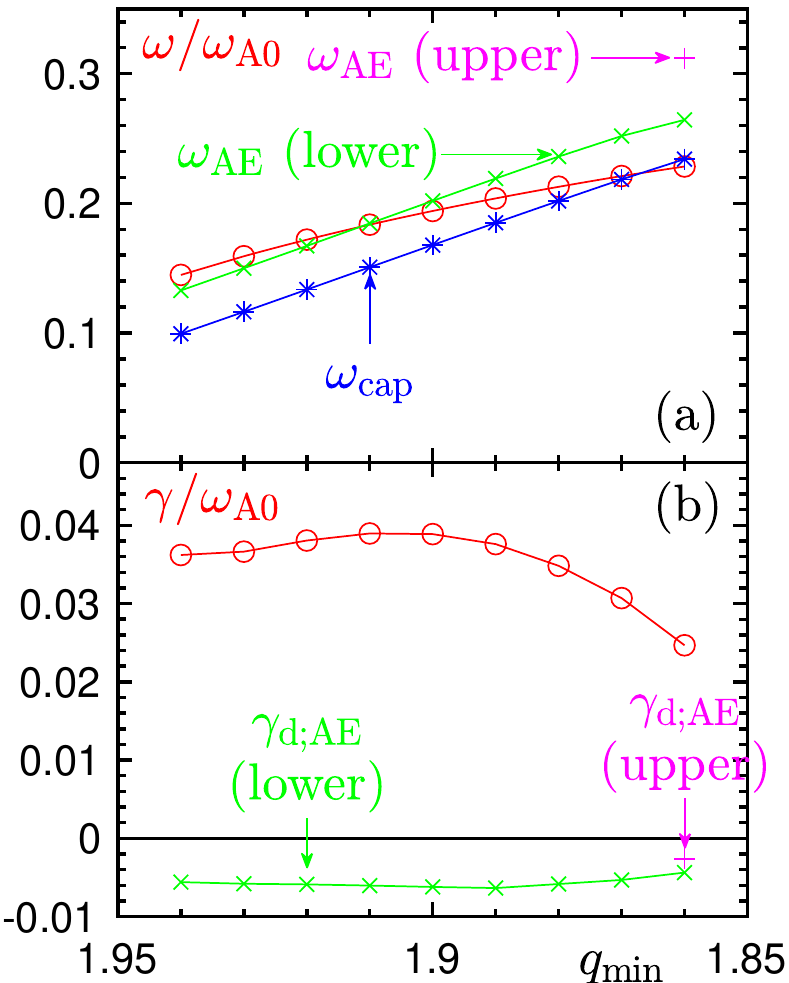}
\caption{Frame (a): the mode frequency $\omega$ (red circles), lower branch (upward sweeping) RSAE frequency in MHD limit $\omega_{\rm AE}$ (green ``x'' markers), and the corresponding continuum accumulation point frequency $\omega_{\rm cap}$ (blue asterisks) as functions of $q_{\rm min}$, which is reported decreasingly to emulate the equilibrium temporal evolution.
Frame (b): the mode linear growth rate $\gamma$ (red circles) and damping rate in MHD limit $\gamma_{\rm d;AE}$ (green ``x'' markers).
For the equilibrium with $q_{\rm min}=1.86$, the upper branch RSAE frequency and damping rate in MHD limit are indicated as magenta crosses.}
\label{fig:linear}
\end{figure}

\begin{figure}
\includegraphics{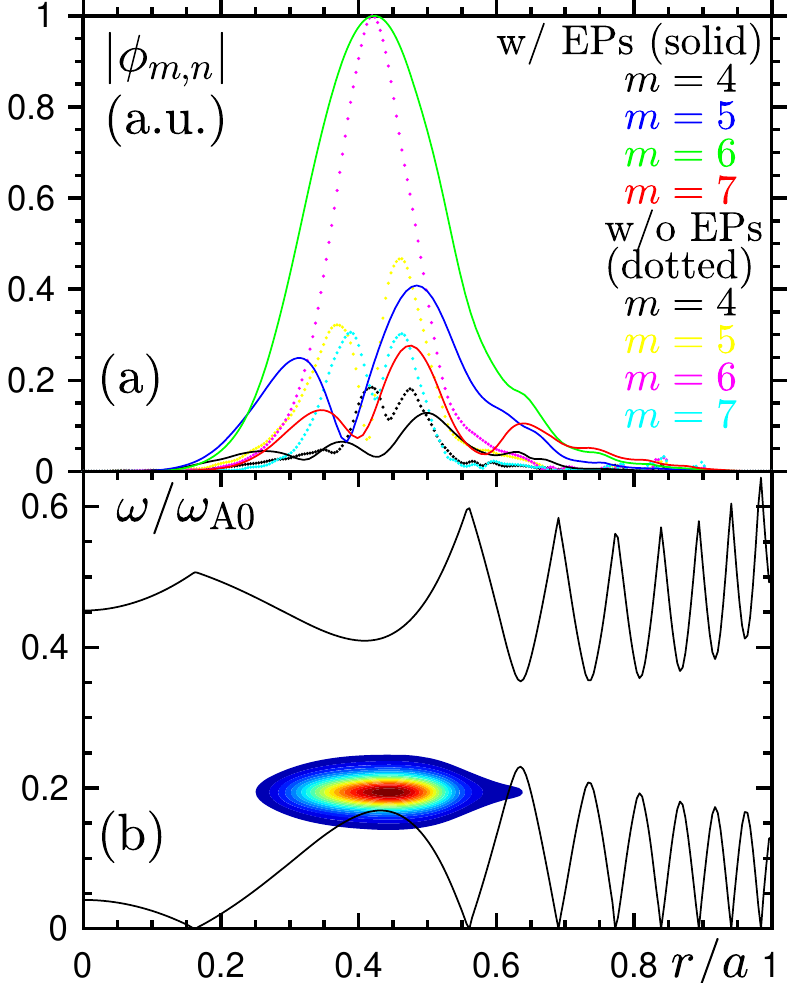}
\caption{For the reference case with $q_{\rm min}=1.90$, several dominant Fourier harmonics of the electrostatic potential $\phi_{m,n}$ in linear stage (a), and the corresponding integrated intensity of FFT spectra $\vert\hat\phi(r,\omega)\vert^2$ (b).
The dotted curves in frame (a) correspond to RSAE radial structure in MHD limit for comparison.
The overlaid black curves in the frame (b) are the ideal MHD SAW continua.}
\label{fig:phi90}
\end{figure}

In the simulations with only varying equilibrium profile controlled by $q_{\rm min}$, we first focus on the RSAE spectra and radial structures excited by non-perturbative EPs.
An overview of fluctuation frequencies and linear growth rates is presented in figure~\ref{fig:linear};
whilst figure~\ref{fig:phi90}
shows radial mode structure for the case with $q_{\rm min}=1.90$ as a reference one for discussion.
We note that the $m=6$ poloidal harmonic is dominant in all cases, while the coupling with the adjacent $m=5$ one intensifies with decreasing $q_{\rm min}$, along with increasing frequency.
Moreover, RSAE spectra and radial structures in the MHD limit are evaluated by antenna excitation \cite{wangtao_pop18} and also reported in figures~\ref{fig:linear} and \ref{fig:phi90} for comparison, where one can observe the significant difference of non-perturbatively EP-induced mode structure and complex frequency shift.
Here, we note that in the underlying equilibrium magnetic fields, RSAE exists as a weakly damped AE due to finite toroidal mode coupling acting as an effective potential well \cite{breizman_pop03}; whereas in the presence of EPs, their non-resonant response extends the potential well and clearly increases radial mode width by their global drive \cite{tsai_pfb93,zonca_pop00,berk_prl01,zonca_pop02}.
Besides, mode damping is mostly induced by finite resistivity (radiative damping) in the MHD limit \cite{matt_pfb92,zonca_pop96,zonca_pop02}; in addition, continuum damping could become important if the fluctuation significantly couples with SAW continuum due to EP-induced frequency shift.
Furthermore, for $q_{\rm min}$ sufficiently close to $(m-1/2)/n$, there also exists an upper branch RSAE dominated by $m=5$ with a frequency below the upper continuum tip \cite{zonca_pop02,breizman_pop03,breizman_pop05}; however, they are not observed to be driven unstable by the sub-Alfv\'enic ($v_{\rm H}/v_{\rm A}<1$) EPs in linear stage, consistent with analytical theory \cite{zonca_pop02}.
Nevertheless, for the case with $q_{\rm min}=1.86$, the upper branch RSAE becomes relevant in nonlinear stage, and is also indicated in figure~\ref{fig:linear} for the sake of completeness.

\begin{figure}
\includegraphics{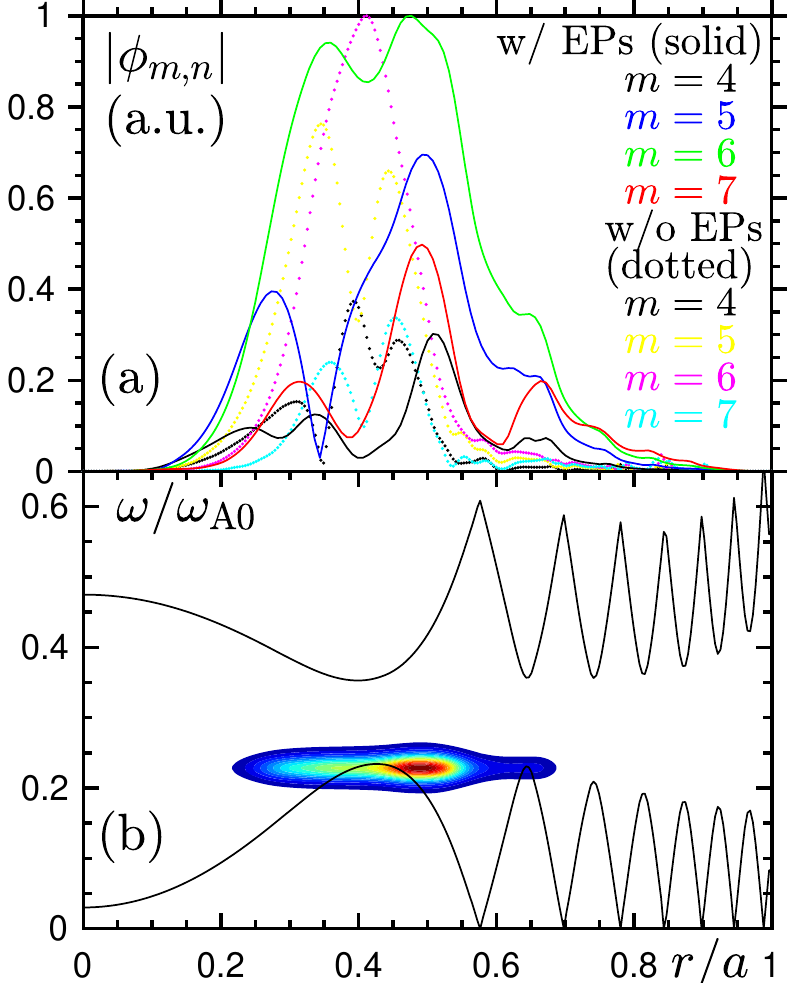}
\caption{Same as figure~\ref{fig:phi90} for the case with $q_{\rm min}=1.86$.}
\label{fig:phi86}
\end{figure}

The impact on radial mode width by non-perturbative EPs is evident in other cases, where the most significant mode structure deformation is found for the case with $q_{\rm min}=1.86$ (see figure~\ref{fig:phi86}).
Its frequency also shows the largest difference with respect to MHD limit; in fact, we note that this mode should be understood as an EPM, as it resides inside the continuum rather than the toroidicity-induced frequency gap.
However, there is no sharp distinction between AE and EPM in this scenario \cite{zonca_pop14a,zonca_pop14b,zonca_njp15,chen_rmp16}, since all the fluctuations are strongly driven by non-perturbative EPs.
They are treated similarly in the following discussions from the perspective of non-perturbative SAW-EP interplay.

\subsection{Resonance analysis}
\label{subsec:resonance}

\begin{figure}
\includegraphics{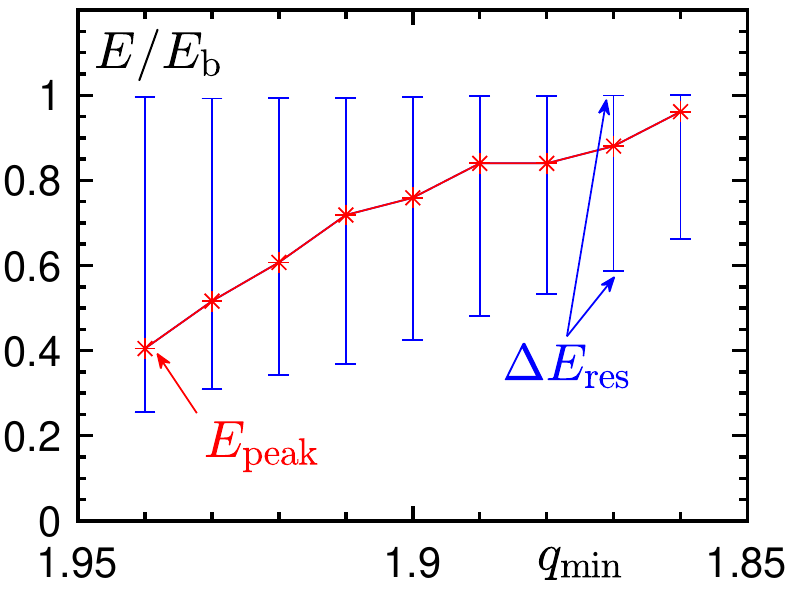}
\caption{Red markers represent the peak of SAW-EP power transfer in EP energy space, $E_{\rm peak}$, by integrating over other phase space variables.
Blue errorbars show approximately the effective resonant energy range, $\Delta E_{\rm res}$, which is estimated as the extent with higher than half of ${\rm Power}(E_{\rm peak})$ in the each case.}
\label{fig:powerE}
\end{figure}

More details underlying the mode excitation, and in particular, the EP-induced frequency shift can be unveiled by wave-particle resonance analysis.
In order to identify the resonant EPs responsible for mode destabilization, we first focus on the phase space structure of SAW-EP power transfer \cite{briguglio_pop14}.
Figure~\ref{fig:powerE}
shows the properties of linear stage power transfer in EP energy space, which is a convenient identifier of the EPs with a narrow pitch angle distribution.
One sees that the peak location $E_{\rm peak}$, corresponding to strongly driving EPs satisfying resonance condition $\omega_{\rm res}\simeq\omega$, shifts to the high energy end with increasing mode frequency.
Here,
\begin{equation}
\omega_{\rm res}=(n\overline{q}-m+\ell)\omega_{\rm t} + n\overline{\omega_{\rm d}}
\label{eq:omega-res}
\end{equation}
is the orbital resonance frequency \cite{briguglio_pop14,zonca_njp15} for co-circulating EPs; $\overline{q}$ is the orbit averaged value of safety factor $q$; $\ell$ is an effective transit resonance harmonic, where $\ell=2$ (first sideband) is found for EPs with $E\simeq E_{\rm peak}$ in all cases.
Note that for the sub-Alfv\'enic beam ions with large orbits, $\ell=\pm1$ is the ``primary'' transit resonance harmonic with most efficient power transfer.
Moreover,
\begin{equation}
\omega_{\rm t}=\frac{2\pi}{\oint {\rm d} \theta/\dot\theta}\simeq \frac{v_\parallel}{qR}
\end{equation}
is the transit frequency;
\begin{equation}
\overline{\omega_{\rm d}} = \oint\frac{\omega_{\rm t}}{2\pi}\left(\dot\zeta/\dot\theta-q\right){\rm d} \theta
\end{equation}
is the toroidal precessional drift frequency, and is generally subdominant to the first term in the right hand side of (\ref{eq:omega-res}) for well circulating EPs.
To the lowest order, the resonance condition with evolving $q$ can be estimated as
\begin{equation}
\omega\simeq\omega_{\rm res}\sim(nq-m+\ell)\frac{\sqrt{E_{\rm peak}/m_{\rm H}}}{qR}\,.
\end{equation}
Thus, $E_{\rm peak}$ necessarily shifts up with increasing $\omega$ and decreasing $q$ for a fixed $\ell$, such that the RSAE upward frequency sweeping corresponds to a similar sweeping in resonant EP phase space distribution as shown in figure~\ref{fig:powerE}.
One could further expect a relay of dominant resonances via higher/lower $\ell$ for higher/lower RSAE frequency; this is confirmed in similar simulations with smaller/larger $q_{\rm min}$ (not reported in this paper).
Nevertheless, power transfer via adjacent $\ell$ becomes notable when $E_{\rm peak}$ approaches the margin of $\Delta E_{\rm res}$ [see figure~\ref{fig:power86}(a) below for an example].
The sweeping in resonant energy and relaying of $\ell$ indicate that RSAE could stay in resonance with substantial EP population during long timescale frequency sweeping; consequently, RSAE-induced EP transport is expected to be global in EP phase space.
The variable resonance condition poses a serious challenge to actively controlling RSAE's excitation and EP confinement degradation.

For the strongly driven RSAEs in the present work, figure~\ref{fig:powerE} also shows that the effective resonant energy range $\Delta E_{\rm res}$ is generally broad, as correlated with the natural frequency extension of a wave with finite growth rate.
Quantitatively, the integrated power transfer can be symbolically written as \cite{briguglio_pop14}
\begin{eqnarray}
{\rm Power} \sim &\int{\rm d}r{\rm d}\mu{\rm d}v_\parallel
\Bigg\langle\frac{e_{\rm H}m_{\rm H}rR}{\omega+{\rm i}\gamma-\omega_{\rm res}}
\nabla f_{\rm H} \cdot {\bf b} \times \nabla \left(\phi-\frac{v_\parallel}{c}\delta A_\parallel \right) \nonumber\\
&\times \left[\left( \frac{e_{\rm H}\mu}{m_{\rm H}}\nabla\log B 
+ \frac{e_{\rm H}v_\parallel^2}{\Omega_{\rm cH}}\kappa\right)\cdot {\bf b}
\times \nabla \phi \right]
\Bigg\rangle_{\theta,\zeta}
\label{eq:power}
\end{eqnarray}
following the EP phase space representation in HMGC, with $\kappa\equiv{\bf b}\cdot\nabla{\bf b}$ the equilibrium magnetic field curvature and ${\bf b}\equiv{\bf B}/B$.
(We refer interested readers to \cite{briguglio_pop14} for more details.)
Here, finite $\gamma$ enters the resonance denominator and broadens $\Delta E_{\rm res}$; in addition, the energy dependence of power transfer [second row in (\ref{eq:power})] also extends $\Delta E_{\rm res}$ to $E\simeq E_{\rm b}$ in all cases, since the power transfer nominator scales as $E^{5/2}$ \cite{chen_prl84} and the underlying distribution function (\ref{eq:df}) roughly scales as $E^{-3/2}$ above $E_{\rm c}$ and up to $E_{\rm b}$.
Thus, the different direction and scales of EP-induced frequency shift can be understood from the asymmetry about $E_{\rm peak}$ within the broad $\Delta E_{\rm res}$; i.e., the shifted fluctuation frequency is weighted by the contribution of all resonant regions so as to maximize wave-EP power transfer.
Furthermore, the variation of $\gamma$, which reflects integrated power transfer, is also consistent with the trend of $\Delta E_{\rm res}$.

\begin{figure}
\includegraphics{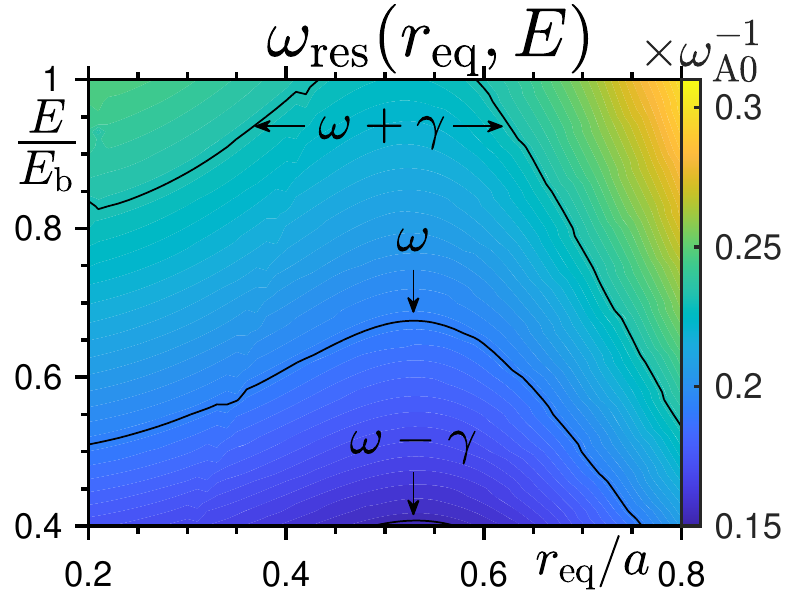}
\caption{For the reference case with $q_{\rm min}=1.90$, contour of EP resonance frequency $\omega_{\rm res}$ in $(r_{\rm eq},E)$ space by following test particle orbits in equilibrium magnetic fields with negligible perturbation.
Here, $r_{\rm eq}$ is a particle's radial coordinate when it crosses the low field side (LFS) equatorial plane (i.e., $\theta=0$) in its equilibrium orbit, and is used as a convenient method to initialize and identify the test particles.
All test particles are characterized by $\alpha_{\rm eq}=\pi/4$ with similar definition.
$n=3$, $m=6$ and $\ell=2$ are used for calculating $\omega_{\rm res}$.
$\omega$ and $\omega\pm\gamma$ are indicated as thin black contour lines for references.}
\label{fig:res90}
\end{figure}

\begin{figure}
\includegraphics{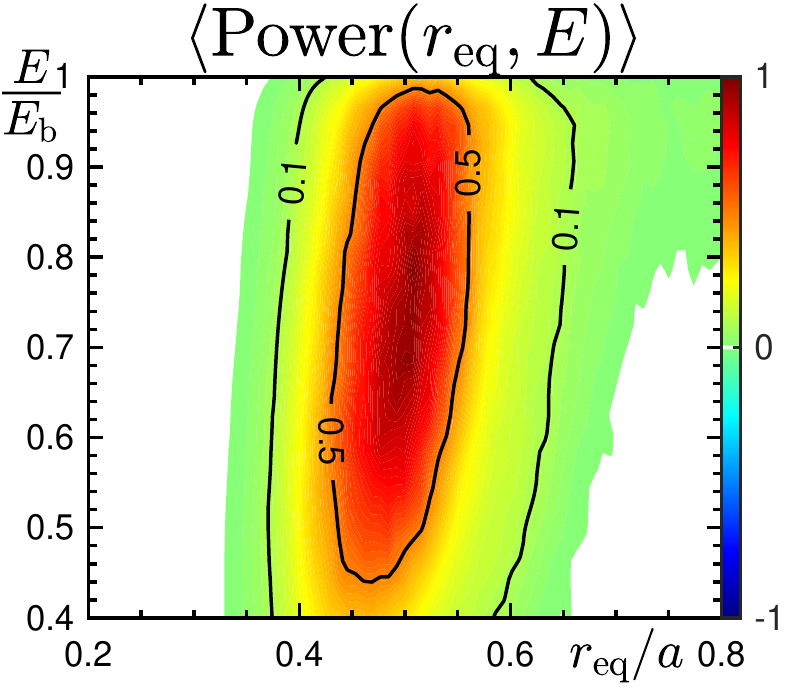}
\caption{For the reference RSAE case with $q_{\rm min}=1.90$, contour of wave-EP power transfer in $(r_{\rm eq},E)$ space by integrating over the pitch angle and averaging over a time window in linear stage.
Here, the positive sign is defined as energy transfer from EPs to the wave; the contour lines with 10\% and 50\% of peak value are explicitly indicated.
That the characteristic power transfer radial scale length is clearly narrower than the radial volume satisfying the resonance condition for effective power transfer $\vert \omega-\omega_{\rm res}\vert \lesssim \gamma$ (cf. figure~\ref{fig:res90}), and is comparable with radial mode width shown figure~\ref{fig:phi90}.
Note that the weak power transfer with larger $r_{\rm eq}$ and $E$ at top right corner is via higher order $\ell=3$.}
\label{fig:power90}
\end{figure}

The above qualitative analysis can be further extended by numerically calculating $\omega_{\rm res}$, where we utilize test particles to represent the phase space region relevant for effective power transfer.
The test particles are evolved in the perturbed electromagnetic fields stored from the self-consistent simulation \cite{briguglio_pop14}; thus, they can also be used as markers to analyze nonlinear dynamics (cf. section~\ref{sec:saturation}).
Figure~\ref{fig:res90}
shows the result of this exercise, where the $(r,E)$ space is mapped by test particles using the injection pitch angle.
One sees that for the sub-Alfv\'enic EPs, the variation of $\omega_{\rm res}$ is quite small in energy space, such that significant power transfer could takes place within a broad phase space range (see figure~\ref{fig:power90}).
Note that the broad $\Delta E_{\rm res}$ is a peculiar feature for the present scenario with $\gamma/\omega_{\rm t}\sim0.15-0.35$ (for relevant phase space range), due to large density (strong drive) but low energy (sub-Alfv\'enic) P-NBs.
Furthermore, nearly the same linear $\omega_{\rm res}$ map is obtained for other cases with small inter-case equilibrium difference; thus, the inter-case shift of $E_{\rm peak}$ and $\Delta E_{\rm res}$ can be appreciated from figure~\ref{fig:res90} by fitting different mode frequencies.
Here, as indicated above, for low frequency cases shown in figure~\ref{fig:powerE}, resonances via multiple $\ell$ contribute to power transfer and consequently, the broad $\Delta E_{\rm res}$.
In other words, although the phase space resonance islands with different $\ell$ do not overlap, the corresponding linear power transfer is nevertheless, not well isolated in phase space, due to the finite $\gamma/\omega_{\rm t}$.

On radial resonance structure, figure~\ref{fig:res90} shows that $\omega_{\rm res}$ is weakly varying radially for a fixed $E$, due to the weak shear dominating the radial derivative of $\omega_{\rm res}$ in (\ref{eq:omega-res}).
Thus, the radial scale length of power transfer is regulated by finite mode width, as compared in figure~\ref{fig:power90}.
The underlying radial resonance structure suggests that nonlinear saturation of the fluctuations will be dominated by the radial decoupling mechanism \cite{vlad_nf13,briguglio_pop14,zonca_njp15,chen_rmp16,wangxin_pop16,vlad_njp16,briguglio_nf17,wangtao_pop19}.
That is, the fluctuation-induced radial EP transport is expected to be comparable with radial mode width, i.e., on meso- or macro-scales, where plasma non-uniformity is crucial to describe the non-perturbative SAW-EP dynamics.
The saturation dynamics is investigated in section~\ref{sec:saturation}.

\section{Nonlinear saturation}
\label{sec:saturation}

\begin{figure}
\includegraphics{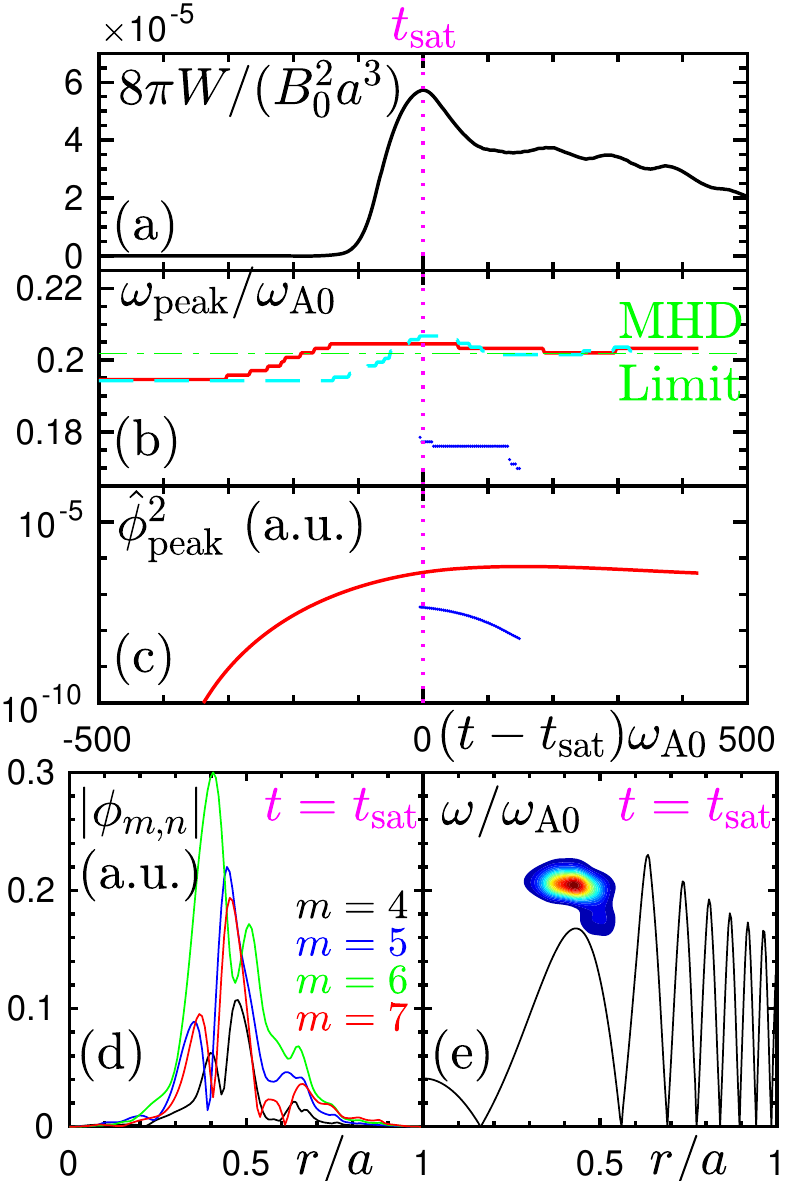}
\caption{Frame (a): for the reference case with $q_{\rm min}=1.90$, the time evolution of volume integrated total (magnetic plus kinetic) energy $W$ as a representative of overall fluctuation amplitude.
Frame (b): time evolution of mode frequency $\omega$ for the primary branch (red solid curve) and a subdominant branch (blue dots), the MHD limit RSAE frequency is indicated as a green dash-dotted horizontal line.
Both series refer to local peak values in $(r,\omega)$ plane of the Fourier transformed intensity $\hat\phi^2(r,\omega)$, using FFT time window $\Delta t_{\rm window}=300~\omega_{\rm A0}^{-1}$.
Similar analysis is performed with $\Delta t_{\rm window}=100~\omega_{\rm A0}^{-1}$ for cross-reference, only the primary branch can be resolved and is shown as cyan dashed curve.
Frame (c): time evolution of peak intensities $\hat\phi_{\rm peak}^2$ of the two branches using $\Delta t_{\rm window}=300~\omega_{\rm A0}^{-1}$ shown in frame (b).
Frames (d) and (e): the radial mode structure and $\hat\phi^2(r,\omega)$ contours at the saturation time.}
\label{fig:mode90}
\end{figure}

We still use the reference RSAE case ($q_{\rm min}=1.90$) for the analysis of nonlinear saturation dynamics in this section, where the relevant spatiotemporal scales are of particular interest.
Meanwhile, the inter-case similarities and differences are summarized at the end of this section.
Figure~\ref{fig:mode90}
presents several macroscopic variables of the fluctuation around the initial saturation time $t_{\rm sat}$, including the volume integrated total fluctuation energy $W\propto\int\phi^2{\rm d}{\bf r}$, mode structures and frequency.
It can be directly observed that when the fluctuation reaches an appreciable amplitude, mode structure undergoes clear deformation from the coherent one in the linear stage; in particular, a splitting in $(r,\omega)$ space is observed.
A primary branch resides at nearly the same radial location with reduced radial mode width and upward chirping frequency; in other words, towards the MHD limit discussed in section~\ref{subsec:spectra}.
In the meantime, a subdominant one gradually emerges, it instead goes slightly downwards in frequency, and clearly decays in intensity.
Here, we note that the actual onset time of primary branch frequency chirping should be shifted forward in time by a large fraction of FFT time window $t_{\rm window}=300~\omega_{\rm A0}^{-1}$, due to the asymmetry of fluctuation amplitude in the FFT temporal series approaching saturation (cf. section~\ref{subsec:MHD}).
For cross-reference, spectral analysis using a smaller value $t_{\rm window}=100~\omega_{\rm A0}^{-1}$ is also reported in figure~\ref{fig:mode90}(b).
As expected, it shows similar chirping range and onset time closer to $t_{\rm sat}$ (but is unable to clearly resolve the subdominant branch).
Thus, it is demonstrated that the nonlinear saturation and frequency chirping occur in the same timescale $\tau_{\rm NL}\sim\Or(\gamma^{-1})$, and are crucially related with the non-perturbative EP dynamics \cite{zonca_nf05,zonca_ppcf15,zonca_njp15,chen_rmp16}.

\begin{figure}
\includegraphics{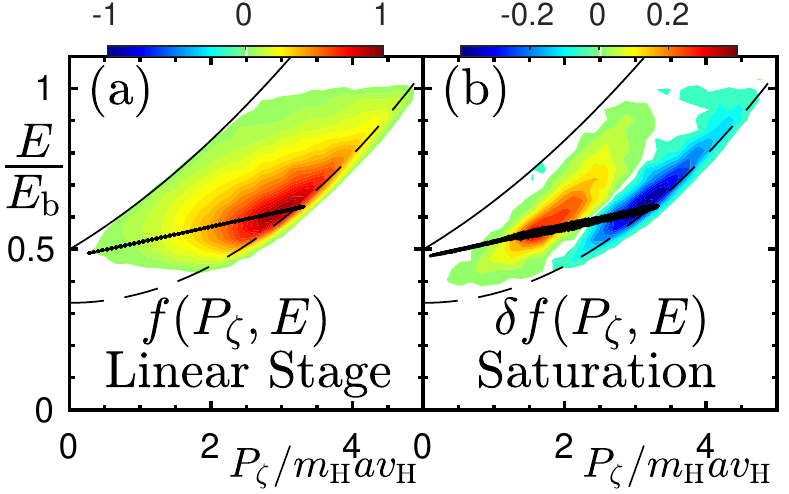}
\caption{For the reference case with $q_{\rm min}=1.90$, normalized EP distribution $f(P_\zeta,E)$ averaged in linear stage (a) and redistribution $\delta f(P_\zeta,E)$ at saturation time (b).
Here, a single value of $\mu B_0/E_{\rm b}=0.336$ is used.
In both frames, the magnetic axis (thin dashed parabola) and LFS boundary (thin solid parabola) are indicated from equilibrium flux surfaces and EP orbits; few EPs fall below the dashed curve are on stagnation orbits \cite{book-white}.
The small black dots in both frames show the test particle phase space coordinates used in figure~\ref{fig:kp90} at the respective time.
Note that the dots remain at the essentially the same line in nonlinear stage, reflecting the conservation of $C\equiv E-\omega P_\zeta/n$.}
\label{fig:df90_com}
\end{figure}

\begin{figure}
\includegraphics{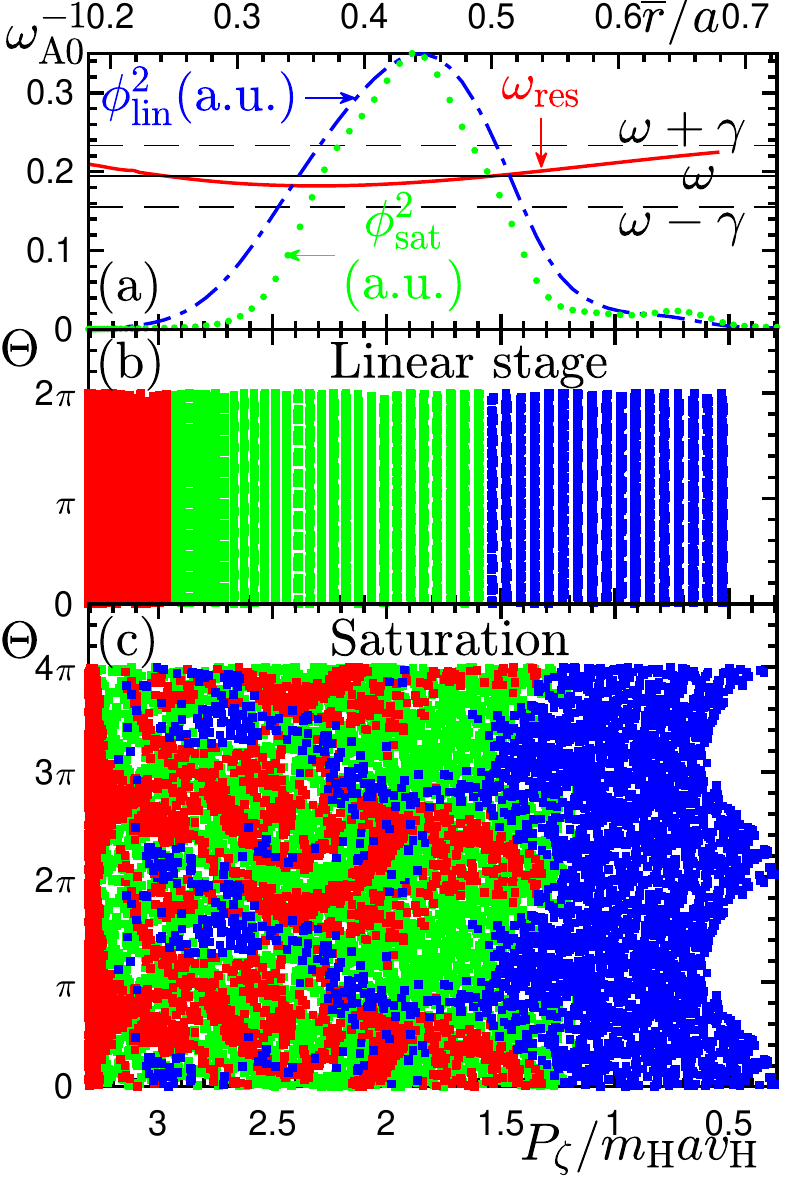}
\caption{Frame (a): for the reference case with $q_{\rm min}=1.90$ in linear stage, the test particles' resonance frequencies $\omega_{\rm res}$ as a function of their toroidal angular momentum $P_\zeta$.
The test particles are initialized with arbitrary spatial distribution and constant $\mu$ and $C$ (cf. figure~\ref{fig:df90_com}).
$\omega$ and $\omega\pm\gamma$ are shown for comparison.
In order to also show the relevant radial scale, mode radial envelope $\phi^2$ at the two points in time plotted in frames (b) and (c) is shown as a function of the test particles' equilibrium orbit averaged radius $\overline r$ (top axis).
Frame (b) and (c): kinetic Poincar\'e mappings in $(P_\zeta,\Theta)$ plane at the linear stage and saturation time, where frame (c) is duplicated in the interval $\Theta\in[2\pi,~4\pi]$ for better visualization.
Each test particle's $P_\zeta$ and the wave-particle phase $\Theta$ at LFS equatorial plane of the last completed orbit are represented as a marker in the Poincar\'e map.
The color scales refer to the initial $P_\zeta$ with respect to a linear resonant value [cf. frames (a) and (b)].
Note that the primary branch mode frequency is used in calculating $\Theta$.}
\label{fig:kp90}
\end{figure}

\begin{figure}
\includegraphics{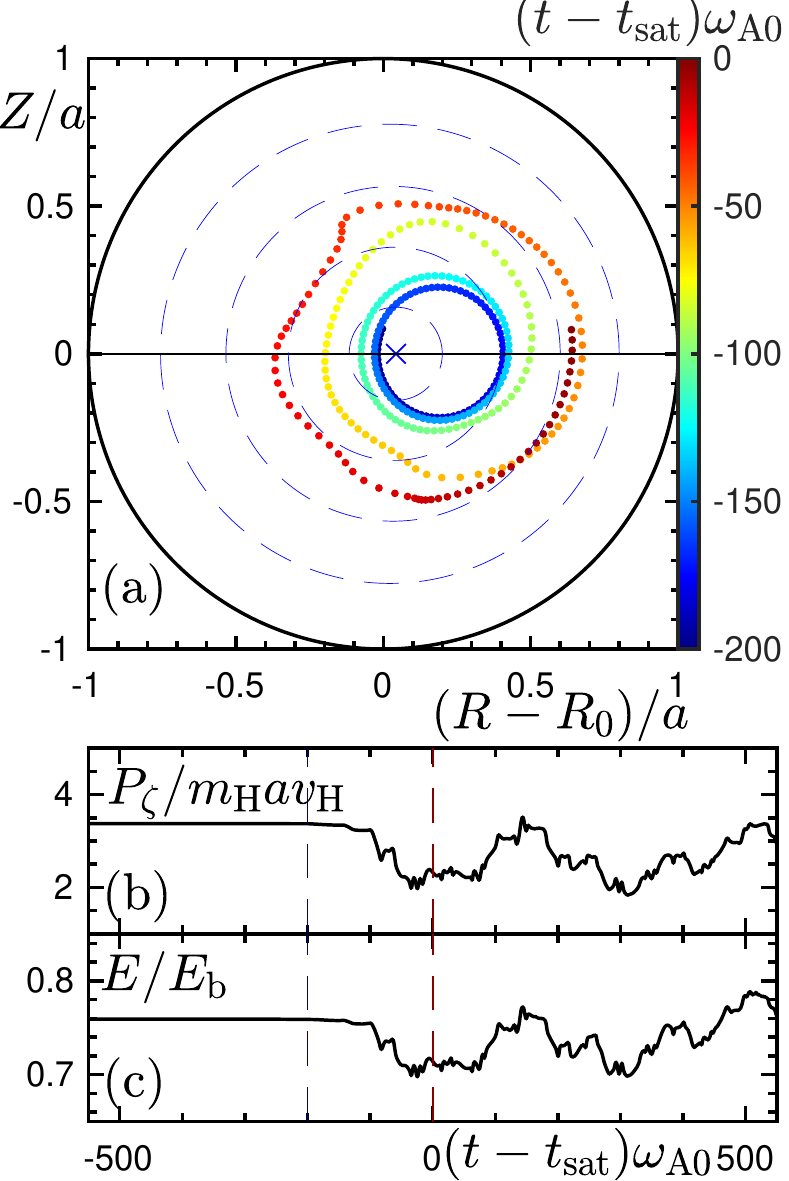}
\caption{Frame (a): for the reference case with $q_{\rm min}=1.90$, a test particle trajectory (colored dots) in poloidal cross section $(R,Z)$ during saturation stage.
The particle is initialized with $r_{\rm eq}/a=0.4$, $E=E_{\rm peak}$, and $\alpha_{\rm eq}=\pi/4$.
The color scale stands for points in time along the orbit.
Blue ``x'' marker and thin dashed lines indicate the magnetic axis and several flux surfaces, respectively.
Frames (b) and (c): the time evolutions of the particle's $P_\zeta$ and $E$ values for reference, where one can note the conservation of $C$.
The plotted range in frame (a) is indicted as vertical dashed lines.}
\label{fig:orbit90}
\end{figure}

As noted above, EP transport should be analyzed in phase space to fully capture its self-consistent interplay with the fluctuations.
Following the constancy of $\mu$, we first examine EP transport in $(P_\zeta,E)$ space, with $P_\zeta=m_{\rm H}Rv_\parallel+e_{\rm H}R_0(\psi-\psi_0)/c$
\footnote{Note that in HMGC, $\psi$ is defined by the form of magnetic field ${\bf B}\equiv B_0R_0\nabla\zeta+R_0\nabla\psi\times\nabla\zeta$; it peaks on the magnetic axis and vanishes on the edge.}
the toroidal canonical angular momentum.
Both $P_\zeta$ and $E$ are orbit invariants in equilibrium magnetic field; thus, the effective perturbation to EP distribution by finite amplitude fluctuation can be readily illustrated by this method, as shown in figure~\ref{fig:df90_com}
for a ``slice'' of EP distribution with a single $\mu$ value.
Significant and global EP transport takes place from inner- to outer-core region, where the distribution essentially follows lines with constant $C\equiv E-\omega P_\zeta/n$.
Here, $C$ is the extended phase space Hamiltonian and is conserved nonlinearly to the lowest order in frequency expansion \cite{book-ll}.
Further to this, a representative phase space portion with single values of $\mu$ and $C$ is sampled by test particles as indicated in figure~\ref{fig:df90_com} \cite{briguglio_pop14}.
Figure~\ref{fig:kp90}
shows the nonlinear evolution of these resonant EP samples in $(P_\zeta,\Theta)$ plane, with $\Theta\equiv -\omega t+n\zeta-m\theta$ the wave-particle phase.
Here, we note that during saturation stage, resonant EPs are nonlocally convected via a secular variation of $P_\zeta$ within the radial volume of mode structure.
In addition, multiple phase space structures corresponding to instantaneous potential wells of the wave \cite{bernstein_pr57,oneil_pf71} can be observed in figure~\ref{fig:kp90}(c), where EPs could also be transported by continuously trapping and detrapping from the potential wells \cite{zonca_nf05,zonca_njp15,chen_rmp16}.
For an illustration of the temporal evolution, figure~\ref{fig:orbit90}
plots the orbit of a representative resonant test particle in both configuration and phase space.
One can readily see a global scale transport within a few orbital periods \cite{white_pf83,briguglio_pop98}, as well as the nonlinear oscillations in the potential well from $P_\zeta$.

On nonlinear spatiotemporal scales, as noted above, the spatial scale of resonant EP transport is dominated by radial convection comparable with radial mode width.
Moreover, at the saturation time, most resonant EPs have not completed one nonlinear oscillation in $P_\zeta$, where the wave-particle trapping time $\tau_{\rm B}$ can be estimated from figure~\ref{fig:orbit90}.
Thereby, nonlinear saturation timescale is comparable with $\tau_{\rm B}$; correspondingly, the frequency chirping shown in figure~\ref{fig:mode90}(b) is non-adiabatic as $\vert\dot\omega\vert\sim\Or(\tau_{\rm B}^{-2})$.
These characteristics make clear that the fluctuation saturates due to radial decoupling with resonant EPs, whose transport is intrinsically nonlocal \cite{zonca_njp15,chen_rmp16}.
In fact, as anticipated in section~\ref{subsec:resonance}, radial decoupling is suggested by the flat $\omega_{\rm res}$ profile shown in figure~\ref{fig:kp90}(a) (or, more generally, figure~\ref{fig:res90}) induced by weak magnetic shear in the inner-core region.
Since the resonance condition $\omega_{\rm res}\simeq\omega$ can be satisfied over a broad radial range, the resonant EPs will not be brought out of resonance by a radially localized orbit excursion.
Instead, in order to quench EP drive necessary for mode saturation, they need to be decoupled from the nonuniform mode structure by radial transport comparable with $\lambda_\perp$.
Thus, for long wavelength RSAEs in weakly reversed shear plasmas, global-scale EP transport by radial decoupling is generally prevalent.
Note that the general validity of this interpretation also applies to RSAEs driven by magnetically trapped EPs, where the radial resonance structure is quite similar to the present case \cite{wangtao_pop19}.

\begin{figure}
\includegraphics{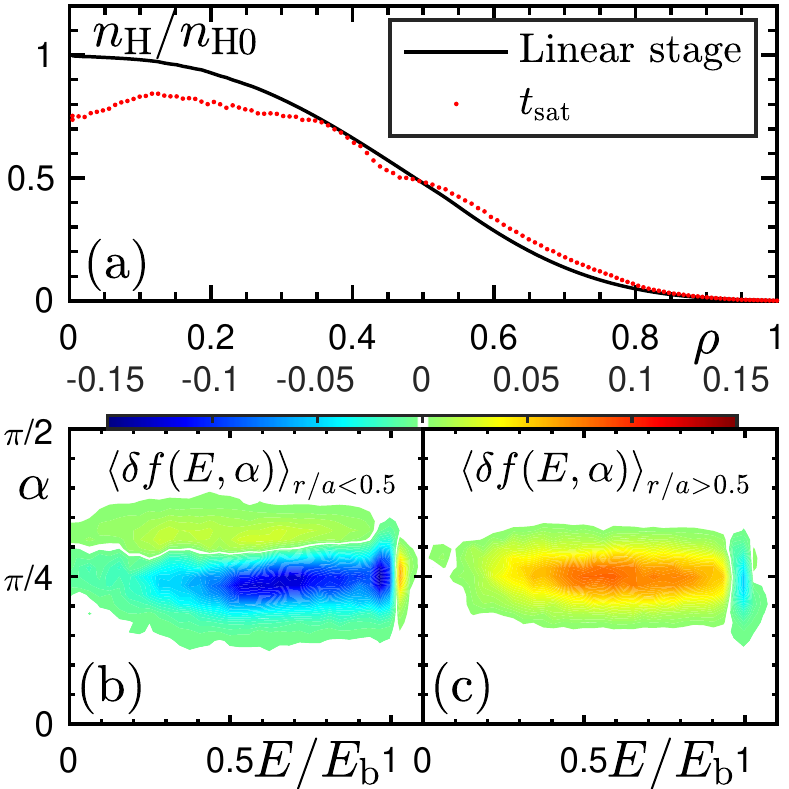}
\caption{For the reference case with $q_{\rm min}=1.90$, EP redistribution at saturation time.
Frame (a) compares EP density profiles in linear stage [thin black curve in figure~\ref{fig:df90_relax}(a)] and saturation time (red dots).
Frames (b) and (c) plot EP velocity space redistribution $\delta f(E,\alpha)$ by integrating over respectively, inner and outer half of radial volume.
Here, $\delta f(E,\alpha)$ is normalized to the linear stage peak value shown in figure~\ref{fig:df90_relax}(b).}
\label{fig:df90}
\end{figure}

Extending the analysis above, the collective EP transport can then be confidently understood following the linear resonance analysis in section~\ref{subsec:resonance}.
Indeed, EP transport occurs in similar spatiotemporal scales for other resonant phase space portions, and results in global EP profile relaxation of the integrated distribution, as shown in figure~\ref{fig:df90}.
In configuration space, EP density profile exhibits a global distortion within radial volume $r/a \lesssim 0.8$, where a profile flattening close to mode peak can be readily seen.
It can be observed that the radial range of density profile relaxation appears to be broader than the radial mode width, due to the fact that resonant EPs generally have large magnetically drift orbit widths (see figure~\ref{fig:orbit90}) \cite{zonca_njp15,chen_rmp16}.
Concurrently, in velocity space, figures~\ref{fig:df90}(b) and (c) evidence a coherent EP flux over approximately $\Delta E_{\rm res}$, analogous to and connected with the configuration space transport via the conservation of $C$, and resulting in a broadening in pitch angle distribution due to conserved $\mu$.

Going back to the frequency spectrum shown in figure~\ref{fig:mode90}, the splitting of mode structure can be qualitatively interpreted in terms of global EP transport as well as radial decoupling from the non-perturbatively driven fluctuation.
Due to nonuniform radial distribution, a finite amplitude fluctuation induces a net-outward flux of resonant EPs, with decreasing $E$ and $v_\parallel$, i.e., decreasing $\omega_{\rm res}$.
If the underlying EP transport is convective and macroscopic, characteristic of non-perturbative SAW-EP interaction in non-adiabatic regime, the fluctuation tends to modify its radial structure and frequency, so as to stay resonant, i.e., phase locked ($\vert\tau_{\rm NL}\dot\Theta\vert,~\vert\tau_{\rm NL}^2\ddot\Theta\vert\ll1$) with the resonant EP flux; such that the corresponding power transfer is maximized \cite{zonca_nf05,vlad_nf13,zonca_ppcf15,zonca_njp15,chen_rmp16}.
Indeed, the downward chirping branch arises from phase locking with resonant EP convection (dubbed as the convective branch in the following).
Note that although the resonance condition is maintained as much as possible by the fluctuation shifting its mode structure and frequency, the convective branch is nevertheless, not strongly driven in the present case, since EP drive intensity associated with its phase space profile gradient is significantly reduced simultaneously.
Meanwhile, it suffers from stronger continuum damping away from the shear reversal radius \cite{zonca_iaea02,zonca_pop02}.
Thus, although the convective branch can be (barely) recognized in figure~\ref{fig:mode90}, it decays in intensity due to both reduced drive and enhanced damping.
On the other hand, accompanied by global EP transport, the non-perturbative effect as shaping the radial mode structure and shifting the mode frequency weakens.
Therefore, there generally exists another branch, which tends to relax to the weakly damped AE state (dubbed as the relaxation branch), and dominates over the convective one for the present case.
The two branches are separated in saturation stage due to radial SAW-EP decoupling.
Experimentally, the existence of the two branches is general for non-perturbatively driven SAW fluctuations \cite{kusama_nf99,shinohara_nf01,shinohara_ppcf04,pinches_ppcf04b,sharapov_nf05,sharapov_nf06,yuliming_prl20}; in particular, it is commonly revealed by fast, non-adiabatic frequency chirping in short timescale, similar to the bursty ``fishbone'' oscillations \cite{mcguire_prl83,chen_prl84}.
Whilst in long timescale observations, the fluctuation either decays significantly until the onset of next burst, where the convective branch transiently dominates in short timescale before the intervention of enhanced damping; or relaxes to the behavior expected in the weakly damped MHD limit with finite amplitude.
Furthermore, we note that the origin of the two branches is conceptually similar to beam and plasma roots in 1D uniform beam-plasma system \cite{oneil_pf68}.
However, diverse behavior is observed for SAW-EP interaction in tokamaks, due to the important role played by magnetic geometry, plasma non-uniformity and non-perturbative EP effect, which govern the relative intensity of two branches via the local and nonlocal EP drive as well as continuum damping.

\begin{figure}
\includegraphics{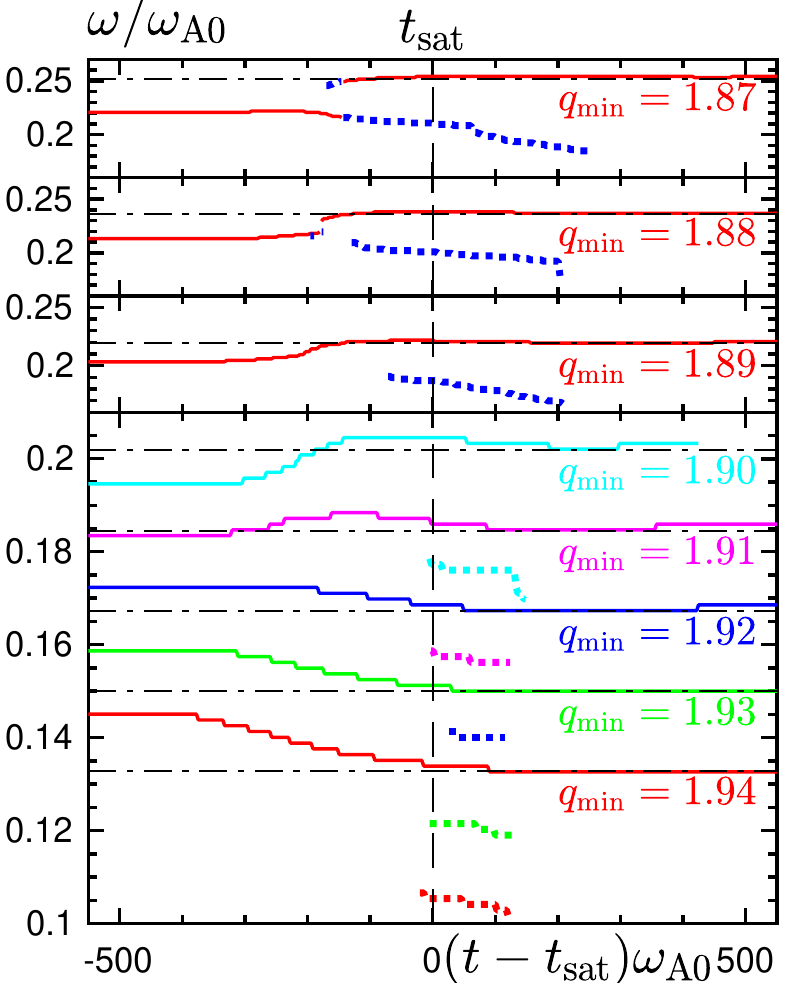}
\caption{For all RSAE cases considered in the present work, i.e., $q_{\rm min}=1.94-1.87$, the time evolution of the mode frequency $\omega$ using similar spectral analysis to that performed in figure~\ref{fig:mode90}.
The solid curves represent the primary branches with largest instantaneous intensity, the squares show subdominant branches, and gray horizontal dash-dotted lines indicate the respective RSAE frequencies in the MHD limit.
The cases with $q_{\rm min}=1.89-1.87$ are plotted in separate frames to avoid confusing overlaps.
Note that the horizontal axis is shifted by the initial saturation time in each case.}
\label{fig:omega}
\end{figure}

\begin{figure}
\includegraphics{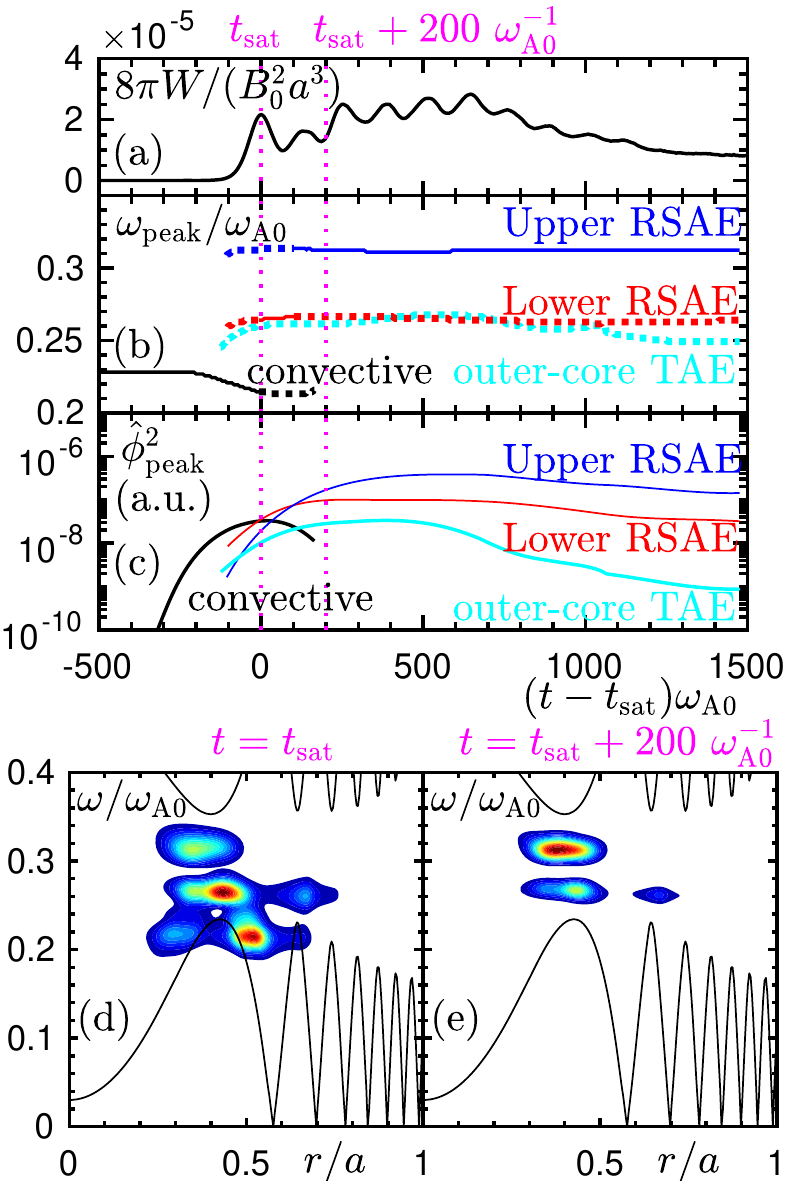}
\caption{Spectral analysis similar to figure~\ref{fig:mode90} for the EPM case with $q_{\rm min}=1.86$.
Here, the plotted time range is extended to post saturation stage for later discussions.
In frame (b), the branches with largest instantaneous intensity are plotted as solid curves, and as squares when they are subdominant.
The four plotted branches are identified by their characteristic radial structure and frequency.}
\label{fig:mode86}
\end{figure}

\begin{figure}
\includegraphics{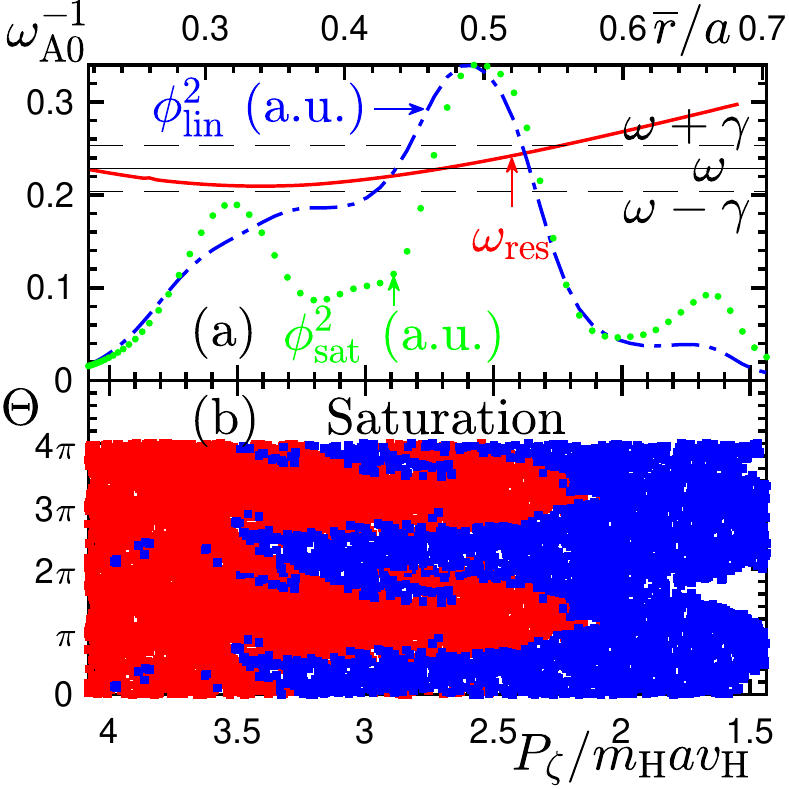}
\caption{For the EPM case with $q_{\rm min}=1.86$, kinetic Poincar\'e map of EP phase space samples, analogous to figure~\ref{fig:kp90}.
Here, the test particles are initialized with constant $\mu$ and $C$ calculated from the following ``reference'' coordinates: $r_{\rm eq}/a=0.5$, $E/E_{\rm b}=0.95$, $\alpha_{\rm eq}/a=\pi/4$.
Note that the plot for linear stage is omitted for simplicity.}
\label{fig:kp86}
\end{figure}

Confidence in the above analysis can be gained by extending the spectral analysis to other cases considered in the present work.
Figures~\ref{fig:omega}
and \ref{fig:mode86}
show similar mode frequency splitting in all cases, illuminating the dominant role played by non-perturbative EPs in the non-adiabatic regime.
In particular, for the case with $q_{\rm min}=1.86$ where the fluctuation exists as EPM wavepackets with strongest non-perturbative EP effect, a convective amplification process prior to saturation can be identified in figures~\ref{fig:mode86}(b) and (c).
Moreover, the phase locking condition for the onset of convective amplification is illustrated in figure~\ref{fig:kp86}
by the Hamiltonian mapping technique.
The ubiquitous resonance detuning process \cite{chirikov_pr79,book-ll} as wave-particle phase shift is minimized, and resonant EPs maintain predominantly near constant phase during radial convective transport.
Finally, anticipating the analysis for the post saturation dynamics in section~\ref{sec:post-saturation}, we note that in all cases, the fluctuation is dominated by weakly damped AEs after saturation, as shown in figures~\ref{fig:mode90}, \ref{fig:omega}, \ref{fig:mode86}; the convective branches are eventually attenuated by enhanced damping.
It suggests that for non-perturbatively driven RSAEs, the relaxation branch as standing wave inside the potential well is preferred over the convective one, due to spatially nonuniform damping and the fact that global EP drive is generally not strong enough to trigger an avalanche process \cite{zonca_iaea99,zonca_nf05,zonca_ppcf15,zonca_njp15}.

\begin{figure}
\includegraphics{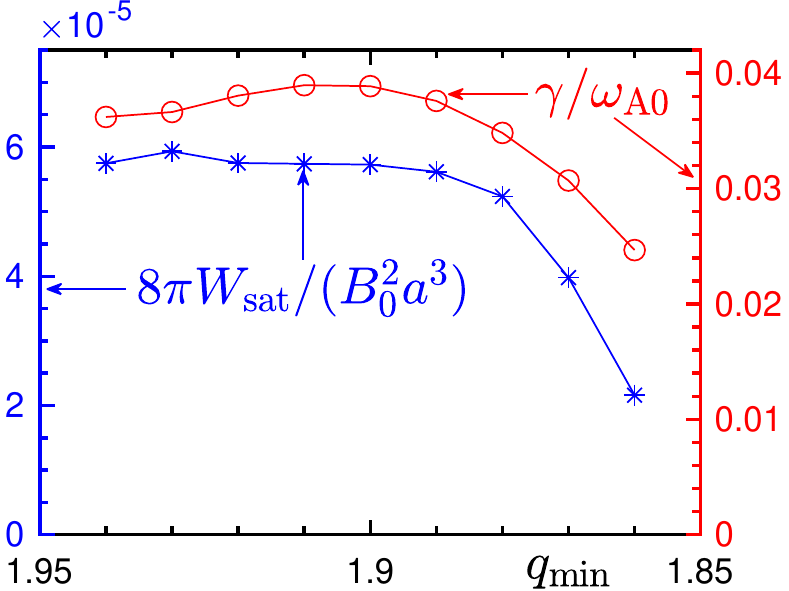}
\caption{For all cases considered in the present work, the fluctuation energy at initial saturation time $W_{\rm sat}$ (blue asterisks, left axis) as a function of $q_{\rm min}$.
Note that the volume integrated energy is preferred over the peak fluctuation amplitude in inter-case comparison, since different radial mode structures are involved.
In addition, their linear growth rates $\gamma$ (red circles, right axis) are reported for reference.}
\label{fig:satamp}
\end{figure}

Similar spatiotemporal scales of EP radial transport are observed in all cases, consistent with their radial resonance structures analyzed and anticipated in section~\ref{subsec:resonance}, as well as the non-perturbative SAW-EP interplay and non-adiabatic frequency chirping illustrated above.
However, collective EP transport has different features at saturation stage, since it is also related with $\Delta E_{\rm res}$ and the fluctuation amplitude responsible for overall transport intensity.
For reference, figure~\ref{fig:satamp}
shows the initial saturation fluctuation energy $W_{\rm sat}$ in qualitative agreement with the trend of $\gamma$.
In accordance with the discussions in this section, EP profile relaxation is similar to the reference case for most low frequency RSAEs; whereas it is clearly weaker in intensity and restricted in high energy extent when approaching TAE frequency, especially for the EPM with most stringent selection of resonant EPs.
However, it is important to note that such inter-case variation is correlated with the variable $v_\parallel/v_{\rm A}\lesssim 1/3$ in the present simulations (cf. section~\ref{subsec:EP}); where the reduction of $\Delta E_{\rm res}$, and consequently, $W_{\rm sat}$, is mostly caused by the transition of dominant resonance harmonic discussed in section~\ref{subsec:resonance}.
In more general cases with a somewhat larger $v_\parallel/v_{\rm A}$ value, RSAE is expected to maintain similar amplitudes during transition to TAE \cite{vanzeeland_pop07}.
Furthermore, it should also be noted that the saturation amplitude does not necessarily represent the post-saturation fluctuation and EP dynamics due to the importance of convective amplification.
More details in post saturation nonlinear dynamics are investigated in section~\ref{sec:post-saturation}.

\section{Post saturation dynamics}
\label{sec:post-saturation}

We focus on the time evolution of fluctuation amplitude and EP confinement properties in this section.
As shown in figures~\ref{fig:mode90}(a) and \ref{fig:mode86}(a), after initial saturation, the amplitude oscillates with a roughly constant period comparable with $\tau_{\rm B}$.
This is consistent with wave-particle trapping in nonlinear stage, where the wave-trapped EPs exchange power back and forth with the wave during oscillation in the potential troughs \cite{oneil_pf71}.
Thus, it is more important to investigate the general decay or amplification in the amplitude, which directly correlates with EP transport intensity.
Here, we note that all RSAE cases are dominated by decay; on the contrary, the EPM with $q_{\rm min}=1.86$ shows clearly enhanced amplitude up to twice of $W_{\rm sat}$ [see figure~\ref{fig:mode86}(a)].
They are discussed separately in the following.

\begin{figure}
\includegraphics{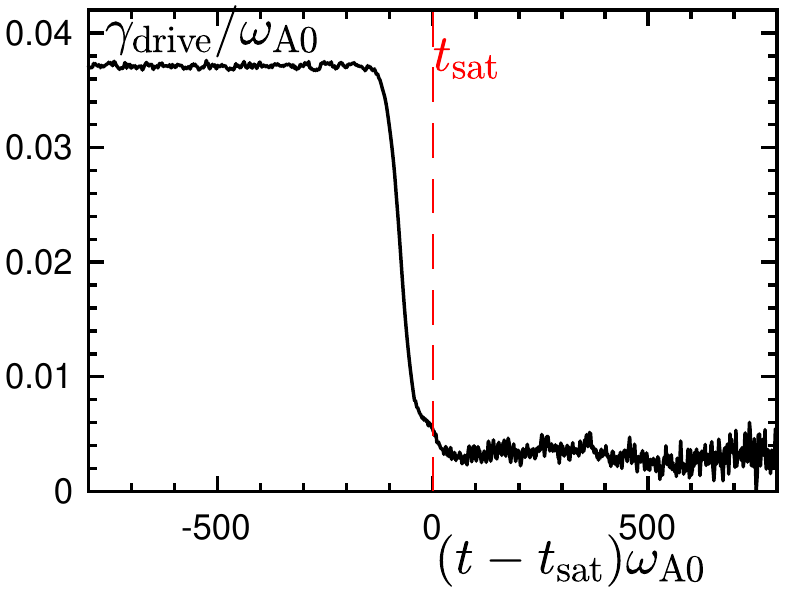}
\caption{For the reference case with $q_{\rm min}=1.90$, time evolution of EP driving rate $\gamma_{\rm drive}$.}
\label{fig:drive90}
\end{figure}

For the decay process dominating the post saturation dynamics of RSAEs, it is straightforwardly related with nonlinear reduction of EP drive intensity and the small but finite mode damping reported in figure~\ref{fig:linear}.
Indeed, for the reference case, figure~\ref{fig:drive90}
plots the time evolution of EP driving rate $\gamma_{\rm drive}$, which is estimated by integrating wave-EP power transfer, weighting over the distribution function and normalizing to the instantaneous total fluctuation energy \cite{briguglio_pop14,wangtao_pop18}.
One sees that associated with the significant phase space profile relaxation during saturation, the residual EP drive is insufficient to overcome the intrinsic damping.
Therefore, the amplitude slowly decays, and EP transport has not significantly increased in post saturation stage.
Here, we note that for the EP population considered in this work, its diamagnetic frequency can be estimated as $\omega_{\star\rm H}\sim m\rho_{\rm H}v_{\rm H}/(r_{q_{\rm min}}L_n)\simeq1.9~\omega_{\rm A0}$, i.e., $\omega_{\star\rm H}/\omega\sim\Or(10)$.
Due to relatively low characteristic EP energy, the condition $\omega_{\star\rm H}/\omega\gg1$ for EP drive dominated by density gradient \cite{book-chen88,fu_pfb89a} is not strictly applicable to the present cases; a substantial fraction of EP drive is also provided by velocity space anisotropy, which can be further pinpointed to the sharp pitch angle distribution (\ref{eq:alfa}).
Correspondingly, the nearly vanishing EP drive should be resorted to the significant transport in both configuration and velocity space \cite{briguglio_pop07} as shown in figure~\ref{fig:df90}.
In particular, we note that in post saturation stage, the outer-core TAE is distinguishable but not strongly driven in all cases (cf. figure~\ref{fig:mode86}), despite increased density gradient in the outer-core region.
Thus, the broad effective resonance range and correlated global profile relaxation in EP phase space are crucially responsible for the decay process.
Furthermore, the fact that RSAE exists as a radially localized standing wave restraints it from being convectively carried away by the outward EP flux.
As noted above, the convective branch is rapidly damped by magnetic geometry and plasma non-uniformity, hence only the weakly damped relaxation branch remains.

\begin{figure}
\includegraphics{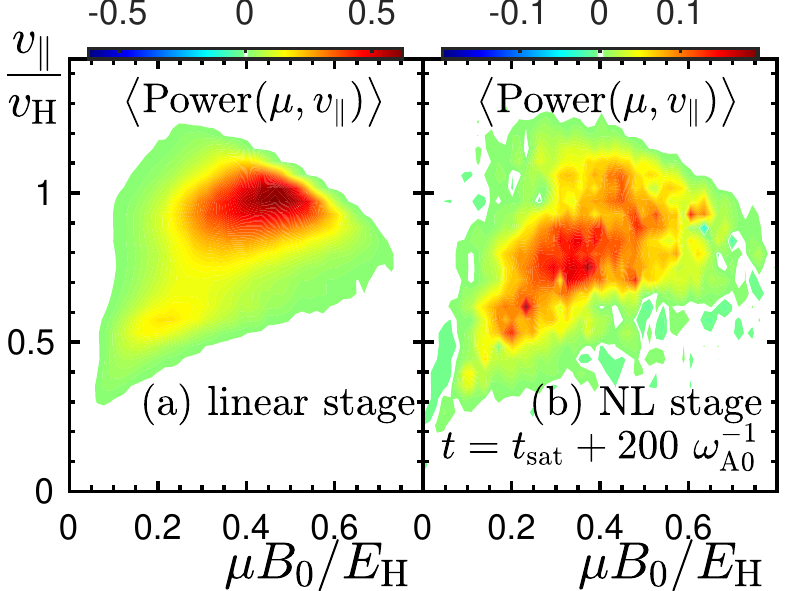}
\caption{For the case with $q_{\rm min}=1.86$, SAW-EP power transfer in $(\mu,v_\parallel)$ space by integrating over plasma volume, where frame (a) is averaged over a time window in linear stage, similar to figure~\ref{fig:power90}, frame (b) corresponds to nonlinear growing stage at the point in time plotted in figure~\ref{fig:mode86}(e).
The absolute value of power transfer is normalized to the instantaneous fluctuation energy [cf. figure~\ref{fig:mode86}(a)].
Note that in linear stage (a), the weak drive at low velocity is via higher order $\ell=3$.}
\label{fig:power86}
\end{figure}

By contrast, EP transport is initially weaker for the EPM case in saturation stage, such that significant free energy is retained in EP phase space distribution.
Via non-adiabatic frequency chirping illustrated in figure~\ref{fig:mode86}, the relaxed fluctuation is able to extend the phase space resonant range and be further amplified, where the highest frequency upper branch RSAE is most strongly driven nonlinearly.
To show this, figure~\ref{fig:power86}
compares SAW-EP power transfer in $(\mu,v_\parallel)$ space during linear and nonlinear growing stage.
We can see that since the phase space resonant region in linear stage is mostly localized in the high energy end, EPs in essentially unaffected phase space region are able to resonate with the relaxed AE and yield significant power transfer in nonlinear stage.
In particular, the upper branch RSAE is strongly driven in post saturation stage due to its large frequency separation from dominant linear fluctuations.
This feature denotes a significant difference with respect to other cases characterized by smaller linear frequency shift and broader phase space resonant range.
Thus, the different nonlinear dynamics are, in fact, produced by the different strength of non-perturbative EP effects, which can be estimated from the scales of linear fluctuation frequency deviation from MHD limit \cite{zonca_pop14a,zonca_pop14b,zonca_njp15,chen_rmp16}.
Ultimately, this deviation is reflected by the corresponding separation of linear and nonlinear resonant ranges in EP phase space, and explains the qualitatively different nonlinear evolutions of fluctuation amplitude.
Here, we emphasize that such nonlinear amplification is induced by non-adiabatic frequency chirping and extension of EP phase space resonant range.
Thus, it is consistent with the theoretical paradigm of convective amplification \cite{zonca_nf05,zonca_ppcf15,zonca_njp15,chen_rmp16}, where the simulations illuminate the fundamental nature of maximizing power transfer in non-perturbative SAW-EP dynamics.
After the residual free energy in EP distribution is exhausted, the fluctuation eventually decays, with significantly enhanced EP transport compared with initial saturation stage.
In the time asymptotic limit of short timescale investigation, i.e., when the fluctuation amplitude decays to a negligibly low value, the overall intensity of EP phase space transport is comparable in all cases.

\section{Conclusion and discussion}
\label{sec:conclusion}

In this paper, via hybrid MHD-gyrokinetic code simulations, we analyze dynamics of reversed shear Alfv\'en eigenmode fluctuation and energetic particles during current ramp-up phase of a conventional tokamak discharge.
The simulations consider time slices in long timescale MHD equilibrium evolution by initializing a series of self-similar $q$ profiles with decreasing $q_{\rm min}$.
Moreover, an anisotropic concentration of fast beam ions is assumed with an idealized model distribution function.
RSAEs are found to be strongly driven by EPs, where the nonlinear dynamics is clearly in non-adiabatic regime dominated by non-perturbative EP response.
In linear stage, EPs induce distortions to mode structure and evident frequency shifts, which can be explained by the global resonance structure.
In particular, the relatively low EP energy restricts linear mode frequency into a narrow band; besides, it also allows a generally broad effective resonance range in EP phase space.
Consistent with the underlying radial resonance structure dominated by weak magnetic shear, the fluctuation radially decouples with resonant EPs during saturation stage; and causes global EP phase space transport, including flattening in density profile, overall slowing-down in energy space, as well as broadening in pitch angle distribution \cite{shapiro_63}.
The nonlinear timescale of saturation dynamics is of the order of wave-particle trapping time, and the spatial scale of EP radial transport is comparable with perpendicular fluctuation wavelength.
Echoing the significant EP transport, the non-perturbatively driven fluctuation chirps non-adiabatically in frequency; by splitting into a downward chirping ``convective'' branch via phase locking with resonant EP convection, and ``relaxation'' branches that tend to evolve into the fluctuation structures obtained in the weakly damped MHD limit.
Associated with the particular SAW continuum structure near shear reversal, the convective branch suffers from enhanced continuum damping; thus, the relaxation branch dominates in post saturation stage.
However, the fluctuation amplitude is still subject to nonlinear EP drive, with implications on overall EP confinement property.
In the present simulations without external source, all RSAEs eventually decay in post saturation stage due to almost quenched EP drive, which is crucially induced by global EP profile relaxation in both configuration and velocity spaces.
Of particular interest, a convective amplification stage is possible, provided that a large frequency separation from relaxed AE state is induced by EPs in linear stage.
This behavior is characteristic of significant non-perturbative EP response, such that the phase space region dominating the resonance structures in nonlinear stage is only weakly disturbed during initial saturation.
In this case, non-adiabatic frequency chirping could broaden effective resonance range in EP phase space, and result in significantly enhanced fluctuation amplitude and EP transport.

This work could be regarded as an attempt to connect fundamental physics of wave-EP resonant interaction with realistic experiments.
On experimental side, the general and specific features of RSAEs strongly driven by high power P-NB ions are analyzed in detail, and the physics underlying diverse observations is illustrated and illuminated.
In general, the present work suggests that during early phase of a discharge with continuous high power input, both linear and nonlinear RSAE dynamics are dominated by EPs, which are usually concentrated in phase space with strong anisotropy.
Indeed, besides the obvious difference of mode frequency and resultant characteristic resonant EP energy, varying $q$ profile across a wide RSAE frequency range does not significantly modify the signature of dominant physics mechanism; all cases clearly show consistent features of non-perturbative SAW-EP interaction.
In other words, ramping-up plasma current produces the foreseeable effect of RSAE frequency sweeping; meanwhile, the fluctuation dynamics are mostly governed by the EP population, which is dominated by high power input and relatively weak dissipation, and is intrinsically unstable to SAW excitation before complete equilibration and thermalization.
Thus, this ``ramping'' scenario generally deviates from marginal stability limit, and might manifest bursty behaviors by repetitive EP accumulation and relaxations \cite{chen_prl84} (see, e.g., \cite{bierwage_nf17a} for recent numerical advances on this topic).

The present work also illustrates many peculiar properties of weakly reversed shear plasmas in typical present-day tokamaks.
For example, radial scale length of EP transport induced by finite amplitude RSAE is expected to be global due to the prevalence of radial decoupling, with the underlying resonance structure dominated by weak magnetic shear.
The general validity of this interpretation is illuminated by \cite{wangtao_pop18,wangtao_pop19} covering a somewhat different parameter regime, where RSAE is driven by the precessional resonance, $\omega\sim n\overline{\omega_{\rm d}}$, of magnetically trapped EPs produced by, e.g., high power on-axis ICRH \cite{cardinali_ppcf20}.
In particular, \cite{wangtao_pop19} shows that even close to marginal stability, RSAE nonlinear dynamics is dominated by radial decoupling due to weak radial dependence of $\overline{\omega_{\rm d}}$ \cite{zonca_njp15,chen_rmp16}.
In the present work, global scale EP profile relaxation takes place due to low P-NB energy with broad effective resonance range in phase space; as well as to the combination of long perpendicular wavelength and large normalized EP orbit width.
Here, we note that most EPs are still confined in the short timescale analysis of this work; i.e., direct RSAE-induced EP loss is negligible with low perturbation amplitude near plasma edge.
However, note also that the redistributed EPs remain energetic in the cooler plasma periphery since $(\Delta E/E)/(\Delta P_\zeta/P_\zeta)\simeq\omega P_\zeta/(nE)\ll1$ for EPs.
Thus, these EPs are prone to be lost by other mechanisms in a more realistic scenario, and hereby, proposing challenges to plasma-wall interaction \cite{dingrui_nf15}.
The ability to effectively scatter EPs makes RSAE a formidable culprit to impacting heating and current drive efficiency as well as first wall power load.
In particular, RSAE instability with flexible frequency is difficult to be controlled from the side of EP phase space engineering, since the effective resonance range sweeps the EP phase space distribution along with sweeping RSAE frequency.
Further analyses are necessary to investigate RSAE-induced EP transport and real time control strategy in practice \cite{garcia-munoz_ppcf19}.

More specifically, the present simulations address the observed multi-timescale frequency chirping and sweeping in experiments \cite{kusama_nf99,shinohara_nf01,shinohara_ppcf04,pinches_ppcf04b,sharapov_nf06}.
It is illustrated that the fast frequency chirping is due to non-perturbatively EP-induced frequency shift and consequent non-adiabatic convection/relaxation, as induced by a large EP phase space concentration with high power input.
The reversed shear configuration always favors the weakly damped relaxation branch via nonuniform continuum damping; therefore, the slow frequency sweeping due to equilibrium evolution is also demonstrated by the time asymptotic limit of short timescale dynamics.
Interestingly, in the case of \cite{kusama_nf99,shinohara_nf01}, after sufficient current ramping, the reversed shear $q$ profile gradually converts to positive shear, whilst similar fast frequency chirping takes place with continuous NBI.
By contrast, in positive shear configuration of the same discharge, without the potential well allowing a weakly damped AE state in adjacent frequency band, nonlinear SAW dynamics is instead dominated by the convective branch \cite{bierwage_nf14,bierwage_pop16b,bierwage_nf17a}; which results into repetitive EPM bursts, and in particular, as the so-called abrupt large-amplitude event.

On the side of fundamental physics, this work demonstrates that the paradigm of non-perturbative SAW-EP interactions in non-adiabatic regime \cite{zonca_njp15,chen_rmp16} is practically important; and investigates the interplay by magnetic geometry, plasma non-uniformity and non-perturbative effects for a case of practical interest.
For the investigated ramping scenario, the adiabatic and perturbative assumption is generally inapplicable, and one must consider the non-adiabatic regime as a general basis.
The simulations illustrate the interconnections among deviation from marginal stability, non-perturbative SAW-EP interplay, non-adiabatic frequency chirping and nonlocal EP transport via radial decoupling, where one can also appreciate their relative importance from inter-case comparison.
In the most representative example where the fluctuation exists as EPM wavepackcts in linear stage, phase locking is shown as a signature for the onset of nonlocal behaviors \cite{zonca_nf05,zonca_njp15,chen_rmp16}.
Indeed, SAW-EP power transfer is maximized by minimizing resonance detuning as well as extending finite interaction time ($\sim\tau_{\rm B}$) and length ($\sim\lambda_\perp$) \cite{zonca_njp15,chen_rmp16}; such that the fluctuation amplitude is convectively amplified, accompanied by nonlocal SAW propagation and non-adiabatic frequency chirping, consistent with various previous theoretical and numerical works \cite{zonca_iaea99,briguglio_pla02,zonca_iaea02,zonca_nf05,briguglio_pop07,vlad_nf09,vlad_nf13,bierwage_nf13,bierwage_nf14,zonca_njp15,chen_rmp16,vlad_njp16,bierwage_nf17a,wangtao_pop19}.
Furthermore, the convective amplification process also extends to the post saturation stage, where the fluctuation is shown to extend phase space resonant range via non-adiabatic frequency chirping, so as to further tap the free energy from EPs.
All the linear and nonlinear dynamics can be conveniently described by the fundamental nature of maximizing wave-EP power transfer \cite{zonca_njp15,chen_rmp16}.
Altogether, the present work demonstrates the importance of taking into account all relevant physics ingredients to properly tackle the crucial and complex issue of EP physics.

Finally, we remind the essential ingredients of the reference scenario adopted by this work, as well as important effects not taken into account in the present analysis.
As discussed above, many features are associated with weak magnetic shear and relatively low EP energy, characteristic of a present-day tokamak configuration.
The reference scenario is expected to be significantly different for future burning plasmas \cite{chen_rmp16}, where the ratio of EP thermal speed to Alfv\'en speed is close to or larger than unity \cite{wangtao_pop18,wangtao_pop19}, and magnetic shear could be strongly reversed due to high bootstrap current fraction (advanced tokamak scenario) \cite{shimada_nf07,chapman_fed11,wanyuanxi_nf17}.
With super-Alfv\'enic EPs, the upper branch RSAE could be strongly excited \cite{kusama_nf99,shinohara_nf01,shinohara_ppcf04}; meanwhile, the SAW spectrum is expected to be much broader with generally smaller normalized EP orbits \cite{gorelenkov_nf14,chen_rmp16,wangtao_pop18,wangtao_pop19}, where a realistic prediction of EP transport must consider the simultaneous presence of multiple unstable waves \cite{white_ppcf10}.
Furthermore, the current ramp-up phase is considered in this work, where the MHD equilibrium is (relatively) quickly evolving, and EP distribution significantly deviates from marginal stability threshold.
On the contrary, for steady state operation in much longer timescale \cite{gong_pst17,gong_nf19}, it is necessary to explicitly include EP source/sink and equilibration effects such as collisions, so as to capture the nonlinear equilibrium evolution of EP distribution; that is, the EP phase space zonal structures \cite{zonca_njp15,chen_rmp16,falessi_pop19a,falessi_njp20}.

\ack

This work is supported by National Key R\&D Program of China under Grant No.~2017YFE0301900.
T.W. acknowledges the support by Shenzhen Clean Energy Research Institute.
This work is also carried out within the framework of the EUROfusion Consortium and receives funding from the EURATOM research and training programme 2019-2020 under Grant Agreement No.~633053 (Project No.~WP19-ER/ENEA-05).
The views and opinions expressed herein do not necessarily reflect those of the European Commission.
The computing resources and the related technical support used for this work have been provided by CRESCO/ENEAGRID High Performance Computing infrastructure and its staff \cite{CRESCO}.

\section*{References}


\providecommand{\newblock}{}

\end{document}